\documentclass[12pt]{article}

\usepackage[utf8]{inputenc}
\usepackage{charter}
\usepackage{fullpage}
\usepackage{hyperref}
\usepackage{graphicx}
\usepackage{amssymb}
\usepackage{lipsum}
\usepackage[sort&compress,numbers]{natbib}
\usepackage{hyperref}
\hypersetup{hidelinks}
\usepackage{wrapfig}
\usepackage[total={6.5in,9in},top=1in,headsep=0.1in,headheight=1in]{geometry}

\newcommand\snowmass{
\begin{center}
  \rule[-0.2in]{\hsize}{0.01in}\\
  \rule{\hsize}{0.01in}\\
  \vskip 0.1in
  Submitted to the Proceedings of the US Community Study\\ 
  on the Future of Particle Physics (Snowmass 2021)\\
  \rule{\hsize}{0.01in}\\
  \rule[+0.2in]{\hsize}{0.01in}\\[-2em]
\end{center}
}
 
\usepackage[firstpage=true]{background}
\backgroundsetup{contents={\parbox{6.5in}{\snowmass}}, scale=1,placement=top,opacity=1,color=black,position={3.25in,1.2in}}

\usepackage{fancyhdr}
\fancypagestyle{plain}{%
  \fancyhf{}%
  \fancyhead[C]{}
  \fancyfoot[C]{\thepage}
}

\fancypagestyle{empty}{%
  \fancyhf{}%
  \fancyhead[C]{{\it Snowmass 2021: Photon counting in visible and near-IR}}
  \fancyfoot[C]{\thepage}
}
\title{Photon counting from the vacuum ultraviolet to the short wavelength infrared using semiconductor and superconducting technologies}
\date{}

\usepackage{authblk}

\author[1]{Jonathan Asaadi}
\author[2]{Dan Baxter}
\author[3]{Karl K. Berggren}
\author[2]{Davide Braga}
\author[4]{Serge A. Charlebois}
\author[5,14]{Clarence Chang}
\author[13]{Angelo Dragone}
\author[2,14]{Alex Drlica-Wagner}
\author[2]{Carlos O. Escobar}
\author[2]{Juan Estrada}
\author[2]{Farah Fahim}
\author[17]{Michael Febbraro}
\author[2]{Guillermo Fernandez Moroni}
\author[6]{Stephen Holland}
\author[7]{Todd Hossbach}
\author[8]{Stewart Koppell}
\author[9]{Christopher Leitz}
\author[12,15]{Agustina Magnoni}
\author[10]{Benjamin A. Mazin}
\author[4]{Jean-François Pratte}
\author[11]{Bernie Rauscher}
\author[12]{Dario Rodrigues}
\author[13]{Lingjia Shen}
\author[14]{Miguel Sofo-Haro}
\author[2]{Javier Tiffenberg}
\author[13]{Joshua Turner}
\author[13]{Lorenzo Rota}
\author[13]{Christopher J. Kenney}
\author[4]{Frédéric Vachon}
\author[5]{Gensheng Wang}

\affil[1]{The University of Texas at Arlington, Texas, USA}
\affil[2]{Fermi National Accelerator Laboratory, Illinois, USA}
\affil[3]{Massachusetts Institute of Technology, Massachusetts, USA }
\affil[4]{Université de Sherbrooke, Québec, Canada}
\affil[5]{Argonne National Laboratory, Illinois, USA }
\affil[6]{Lawrence Berkeley National Laboratory , California, USA}
\affil[7]{Pacific Northwest National Laboratory, Washington, USA}
\affil[8]{Stanford University, California, USA}
\affil[9]{MIT Lincoln Labs,  Massachusetts, USA}
\affil[10]{University of California Santa Barbara, California, USA}
\affil[11]{NASA Goddard Space Flight Center, Maryland, USA}
\affil[12]{Universidad de Buenos Aires, Argentina}
\affil[13]{SLAC National Accelerator Laboratory, California, USA}
\affil[14]{Centro Atomico Bariloche, Argentina}
\affil[15]{University of Chicago, Illinois, USA}
\affil[16]{Laboratorio de Óptica Cuántica, UNIDEF, Argentina}
\affil[17]{Oak Ridge National Laboratory, Tennessee, USA}

\begin{document}

\maketitle 

\begin{abstract}
\noindent In the last decade, several photon counting technologies have been developed opening a new window for experiments in the low photon number  regime. Several ongoing and future projects in HEP benefit from these developments, which will also have a large impact outside HEP. During the next decade there is a clear technological opportunity to fully develop these sensors and produce a large impact in HEP. In this white paper we discuss the need for photon counting technologies in future projects, and present some technological opportunities to address those needs.
\end{abstract}

\def\thefootnote{\fnsymbol{footnote}}
\setcounter{footnote}{0}
%


\begin{center}
	\tableofcontents
\end{center}
\section{Needs for Future Projects}\label{sec:needs}

In this section we describe future projects that will benefit from photon counting detectors in the vacuum ultraviolet, visible, and near-infrared (near-IR), what they need for pixel count, efficiency, timing, etc.

\subsection{Observational Cosmology}\label{sec:cosmology}

\subsubsection{Ground-Based Spectroscopy}

The Cosmic Frontier explores the nature of dark energy, dark matter, and cosmic inflation through the observations of faint stars, galaxies, and residual microwave photons from the Big Bang.
Several novel cosmological facilities for wide-field multi-object spectroscopy were proposed for the Astro2020 decadal review, and are being considered as part of the Snowmass process \citep[e.g.,][]{MegaMapper,MSE19,2019BAAS...51g.126M,2018arXiv180207216D,DMfacilitiesWP}. 
Ground-based spectroscopic observations of faint astronomical sources in the low-signal, low-background regime are currently limited by detector readout noise.
In particular, medium- to high-resolution spectroscopy at shorter wavelengths has low sky-background levels and significant gains can be achieved through reductions in readout noise (${\sim} 0.5$ e- rms/pix) \citep{Drlica-Wagner:2020}.
Sub-electron noise would result in a $\sim 20\%$ increase in survey speed, i.e., allowing a five year survey to achieve its goals in four years. 
Multi-object spectroscopic facilities are required to observe objects of widely varying brightness with the same fixed exposure time. 
Thus, the ability to control readout noise dynamically would allow photon counting when needed for faint sources, but will not waste time on bright sources that are shot-noise dominated.
Another application for single-photon counting comes from observing campaigns (imaging or spectroscopy) that consist of many short-duration exposures become dominated by the readout noise of each exposure \cite{Richmond:2020,Tingay:2020}.
For example, high-cadence observations searching for short duration transients (e.g., fast radio bursts, gravitational wave events, etc.) could benefit from reduced detector readout noise.

Scientific CCDs are the detector technology of choice for current ground-based observatories. 
CCDs provide many desirable properties for astronomical observations including stability, linear response, large dynamic range, high quantum efficiency over a wide wavelength range, and low dark count rates. 
However, current scientific CCDs have a noise floor at $\sim 2$ e$^{-}$ rms/pix.
Ideally, a future photon counting technology for ground-based astronomical observations would retain many of the beneficial properties of conventional CCDs.
In addition, detectors with lower band gap (i.e., germanium CCDs) would enable observations of higher-redshift galaxies.
However, fast readout time is essential for any new technology, since every second spent in readout is a second lost in exposure time.

\subsubsection{Special Considerations for Space}
\label{sec:space:special}


Space affords certain unique observational advantages. Among these are the complete absence of atmospheric ``windows'' and ``seeing'', no electromagnetic interference except that from the observatory itself, and generally superb thermal/mechanical stability. However, to take advantage of these benefits, detectors are required to operate in the harsh space radiation environment. 

The Decadal Survey 2020 \cite{Decadal2020} has established a large space telescope as the highest ranked large project for the next decades. The missions concepts developed  in this area \cite{habex, luvoir} have identified photon counting arrays for
visible and near-IR imaging and spectroscopy as key enabling technologies. Some of the photon counting technologies discussed here are a good match to these missions. 

As an example, the James Webb Space Telescope (JWST) has a nominal mission lifetime of five years, with all expendables provisioned for ten. Because launch was so successful, we can now expect many more than five years of JWST science in part because everything is radiation tolerant. The detectors themselves were tested to withstand a lifetime dose of 5-10~krad-Si. The SIDECAR application specific integrated circuits (ASIC) that control them were tested to far higher levels still.

\begin{figure}
\begin{center}
\includegraphics[width=1\hsize]{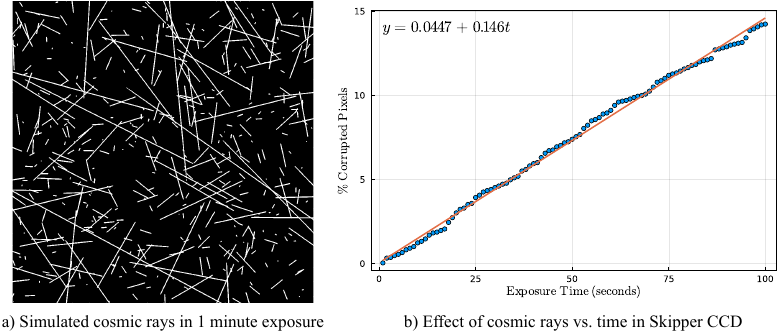}
\end{center}
\caption{Panel a) shows a Monte Carlo simulation of a 250~$\mu$m thick Skipper CCD used at L2. The tracks are long because thick silicon is used to achieve excellent QE in the near-IR. We assumed that a cosmic ray disturbed any pixel that it passed within about 9~$\mu$m of (orthogonal distance) and the 4 nearet-neighbors. Panel b) shows that exposure times should be kept to less than about 1 minute to disturb no more than about 10\% of pixels.}
\label{fig:L2-cosrays}
\end{figure}

Beyond survival, the detectors themselves must still meet demanding performance requirements even though the cosmic ray rate is far higher than on the ground. The integrated ionizing particle rate at L2 is about $5-10~\textrm{ions}~\textrm{cm}^{-2}\textrm{s}^{-1}$. For imaging detectors including CCDs and IR arrays, this places an upper limit on the useful exposure time. Figure~\ref{fig:L2-cosrays} shows the fraction of pixels that cosmic rays would corrupt in a hypothetical p-channel CCD at L2. Total exposure time including readout must be kept shorter than about 1 minute to avoid corrupting more than 10\% of the pixels.

\subsection{Direct Dark Matter search}\label{sec:darkmatter}

Among the most promising detector technologies for the construction of a large multi-kg experiment for probing electron recoils from sub-GeV DM are new generation of silicon Charged Coupled Devices with an ultralow readout noise, so-called ``skipper-CCDs''. These sensors are a new type of photon counters discussed in detail in Section~\ref{sec:skipper}. Skipper-CCDs were designed by the Lawrence Berkeley National Laboratory (LBL) Micro Systems Lab in close collaboration with Fermilab.  In a technological breakthrough in 2017, the SENSEI (``Sub-Electron Noise Skipper-CCD Experimental Instrument'') Collaboration demonstrated the ability to measure precisely the number of 
free electrons in each of the million pixels across the CCD~\cite{skipper2017}. 
SENSEI packaged a small prototype skipper-CCD sensor and took data at Fermilab on the surface and underground, setting world-leading constraints on DM-electron interactions for DM masses in the range of 500~keV to 5~MeV~\cite{sensei2018,sensei2019,Barak:2020fql}.  
Using new, science-grade skipper-CCDs, pathfinder experiments (already funded) based on this technology are planned for the coming years.  Specifically, SENSEI-100 plans to install a $\sim$100~g detector at SNOLAB during 2021, while DAMIC-M plans to install a $\sim$1~kg detector at Modane during 2023. 

The Oscura project has the goal of performing a dark matter search using silicon Charge Coupled Devices with a threshold of two electrons and no background events in a total exposure of 30 kg-yr. The R\&D effort, started in FY20, has focused on the three highest risk technical aspects of the experiment: sensor fabrication, readout electronics, and background reduction. The effort is now moving into a design phase with the plan of being ready to start construction in FY24.

Superconducting nanowire single photon detectors (SNSPDs) (see Sec.\ref{sec:SNSPD}) are also a promising technology for the detection of light dark matter, thanks to their inherent low energy threshold, experimentally demonstrated to be 125 meV \cite{verma2021}, which can enable sensitivity to masses lower than 0.1 MeV when dark matter produces an electron recoil leading to scintillation in a cryogenic semiconductor, such as GaAs~\cite{PhysRevD.96.016026,Vasiukov:2019hwn}. The rate of background dark counts in SNSPDs has been demonstrated to as low as one event per day in a (300 $\mathrm{\mu m)^2}$ pixel, many orders of magnitude lower than in the photosensors used in previous generations of dark matter searches. A key milestone towards a functional dark matter experiment scalable to target masses up to 1 kg is the realization of highly-pixelated SNSPD sensors covering active areas as large as several $\mathrm{cm}^{2}$, while maintaining high time resolution and low noise. Other applications of SNSPDs for the detection of light dark matter include the realization of optical haloscopes based on dielectric cavities \cite{Baryakhtar2018}, searches for dark photons and axions with dish antennas \cite{liu2021broadband} and the direct detection of electron scattering from dark matter in superconductors \cite{Hochberg2019}.

\subsection{Neutrino detection}\label{sec:neutrinos}

Current observation of low energy neutrino are limited by the high energy threshold of available detector technologies. Single photon counting sensors have gained importance for this application due to their ability to access to energy depositions below a few hundred eV's. The CEvNS interaction \cite{FreedmanCEVENS1974}  (Coherent Elastic neutrino Nucleus Scattering ) recently observed \cite{cevns_2017} provides a new mechanism to observe low energy neutrinos (with energy below 50 MeV). Several low energy neutrino sources are available: nuclear reactor (which are the most incense source of neutrino in the earth), spallation neutron source, the sun, radiation sources, etc. This neutrino interaction channel is a good candidate to test the standard model physics at low momentum transfer \cite{fernandez2021WeakMixing} and possible deviations from it \cite{fernandez2021LightMediators} since it can only interact through the weak force with less backgrounds associated with other forces when compared to other elementary particles. The deposited energy from CEvNS is less than a few keV. Only part of this energy is converted into detectable signal in the sensor (ionization, phonons, etc.) and therefore low threshold technologies are needed. Semiconductor and superconducting technologies with eV and sub-eV energy resolution for photon counting capability in the visible and near-IR are natural candidates to reach the necessary resolution for this application.

A large number of outstanding questions remain to the fundemental nature of the neutrino, which can be probed through the use of higher energy ( $\mathcal{O}$(MeV) $< E < \mathcal{O}$(GeV)) neutrino sources (e.g. accelerator and atmospheric neutrinos). The nature of these remaining puzzles break into the distance over which the neutrinos are allowed to propogate before being detected. Thus the future class of experiements are classified as ``short-baseline'' and ``long-baseline'' experiments. 

The next generation long-baseline neutrino experiments aim to answer the questions of the exact ordering of the neutrino mass states, known as the mass hierarchy, as well as the size of the CP-violating phase $\delta$. These, as yet unknown quantities, remain one of the last major pieces of the Standard Model of particle physics and offer the opportunity to answer such fundamental questions as ``what is the origin of the matter/antimatter asymmetry in the universe?'' and ``do we understand the fundamental symmetries of the universe?''. By measuring the asymmetry between appearance of electron neutrinos from a beam of muon neutrinos ($P(\nu_{\mu} \rightarrow \nu_{e}$)) compared to the appearance of electron antineutrinos from a beam of muon antineutrinos and $P(\bar{\nu}_{\mu} \rightarrow \bar{\nu}_{e}$)) as well as the precise measurement of the $\nu_{e}$ energy spectrum measured at the far detector, both the CP violating phase ($\delta_{CP}$) and the mass hierarchy can be measured in the same experiment.

The Short-Baseline Neutrino (SBN) program aims to address the anomalous neutrino results seen by the LSND and MiniBooNE which suggest the possible existence of a eV mass-scale sterile neutrino. However, the experimental landscape is perplexing since a number of other experiments utilizing a range of different neutrino sources which should have been sensitive to such a sterile neutrino have observed only the standard three neutrino oscillations. In this landscape, the conclusive assessment of the experimental hints of sterile neutrinos becomes a very high priority for the field of neutrino physics.

To address both of these areas of neutrino research, large scale noble element time projection chambers (TPC's) \cite{Radeka1974, Rubbia1977} play a central role and offer an opportunity to perform discovery level measurements through the enhancement of their capabilities. In a noble element TPC, particles interact with the medium and deposit their energy into three main channels: heat, ionization, and scintillation light. Depending on the physics of interest, noble element detectors attempt to exploit one or more of these signal components. Liquid Noble TPC's produce ionization electrons and scintillation photons as charged particles traverse the bulk material. An external electric field allows the ionization electrons to drift towards the anode of the detector and be collected on charge sensitive readout or transform energy carried by the charge into a secondary pulse of scintillation light. 

Scintillation light from the interaction produces ultraviolet (UV) photons and is typically detected by separate and independent photosensors located away from the charge sensitive plane and provides the $t_0$. Detecting these UV photons is traditionally accomplished using wavelength shifters to shift the light to the visible wavelength for detection by conventional light detectors like PMTs or SiPMs. This detection technique offers moderate overall detection efficiencies with various levels of complications associated with the use of wavelength shifters and/or UV sensitive devices. 

The combined measurement of the scintillation light and the ionization charge gives the TPC the capability of being a fully active 5D detector with 3D tracking, calorimetric reconstruction capabilities and timing measurements. However, the realization of the full multi-dimensional readout over the full spectrum of produced signals of noble element TPCs remains one of the key instrumentation challenges.

To realize this challenge and unlock the full potential of noble element TPC requires multiple pieces including  large area, high UV efficiency to scintillation light, expanded sensitivity to the full spectrum of light produced, and ideally these capabilities integrated into a single sensor capable of detecting both light and charge. Research into novel photodetection techniques utilizing various new materials is one of the paths which needs to be explored to realize the full capabilites of these experiments.


\subsection{The Material Challenge : Replacing Silicon for Short-Wave Infrared Detection}




As is well known noble elements emit light abundantly in the vacuum ultraviolet (VUV) part of the spectrum \cite{apriledoke2006}, making it natural that the focus of R\&D in the last couple of decades has been on increasing the light collection efficiency of this VUV scintillation using wavelength shifters plus light guides or light traps. Section 2.8  discusses further developments on the direct detection of VUV scintillation light in noble element detectors but now it would be interesting to call attention to the non-VUV part of the spectrum and its potential use as the light signal in such detectors, an application that could bring some advantages in terms of abandoning the use of wavelength shifters as well as bypassing the technological challenges of VUV sensitive photon detectors, not ignoring the fact that going to longer wavelengths has its own technological challenges, as made explicit in the following. 

It has been known since the late 80s that noble gases present atomic emission in the near-infrared (NIR) \cite{lindblom1988,bressiPLA} that would be quenched by collisions  in the condensed phases thus raising the question of NIR emission in liquid and solid arising from excimers. After inconclusive results coming mostly from Bressi, Borghesani, Carugno and collaborators \cite{Borghesanijchemphys}, a revival of the subject was brought forward independently by the Munich group of A.~Ulrich and co-workers \cite{HeindlJINST} and the Novosibirsk group \cite{Bondaretal} with the discovery of NIR scintillation around and beyond 900 nm in liquid argon thus urging the quest for single photon counting devices at wavelenghts larger than 900 nm well into the 1,000 nm silicon brick wall, with the finding of a strong emission in this region when liquid argon is doped with parts per million of xenon \cite{NeumeierEPL}.

Though silicon can absorb photons up to 1110~nm, using silicon above 800~nm requires thick structures with extended and uniform electric field.
Silicon photodiodes therefore come in a variety of configurations optimized to have either frontside photon incidence (i.e. entrance on the high field junction side) or backside incidence (i.e. entrance from the substrate).
These configurations all share the goal of favoring avalanche triggered by electrons \cite{dautet1993photon}.
In all cases, a compromise must be reached between higher sensitivity at the cost of higher dark current which in photon counting detectors translates to higher dark counts rates (larger device volume).
Such devices also have longer drift lengths and thus greater time jitter.

Practical sensitivity in the short-wave infrared (1~to 3~µm) thus requires using other semiconductors such as germanium or small bandgap semiconductor elements.
But dark count rates are greatly increased by the exponential dependence of pair generation rate across that small bandgap.
Competing with silicon with respect to dark count rate is a huge challenge for many reasons.
Firstly, silicon being an indirect bandgap semiconductor sees its band-to-band transition generation rate greatly suppressed (5~orders of magnitude).
Therefore the main generation mechanism is through mid-gap impurity states which depends on the density of those states, not on the doping of the material \cite[Sec.~1.5.4]{Sze2007}.
Secondly, silicon homojunction have the best crystalline structure with minimal defect density.

Infrared avalanche photodiodes have been demonstrated using a variety of small bandgap semiconductors that can achieve single-photon sensitivity.
Of central interest to photon counting are single-photon avalanche diodes (SPAD) based detectors such as silicon photomultipliers (SiPM) and photon-to-digital converters (PDC), also known as digital SiPM\footnote{See section \ref{sec:PDC}}.
All these detectors operate in a metastable regime above the breakdown voltage of the junctions called "Geiger mode" \cite{cova1996avalanche}.

\medskip
In recent years, germanium-on-silicon single-photon avalanche diodes have been demonstrated and used in various applications \cite{lu2011geiger, warburton2013ge,Izhnin2021,Llin2020,Kuzmenko2020,martinez2017single,vines2019high}. 
Although challenging because of the 4.2\% crystal lattice mismatch, being relevant to transistor and optoelectronic technologies, high quality growth of germanium on silicon wafer up to 200~mm is now commercially possible.
This is a relevant issue if one seeks for large sensitive surfaces (SiPM) and integration with advanced electronics (digital SiPM, PDC, CCD, and other imagers).
While germanium also benefits from reduced band-to-band generation rate because of its indirect bandgap (E$_g =$~0.65~eV), the trap-assisted (Shockley-Read-Hall) generation rate in a depletion region is 3~orders of magnitude larger than for silicon.
Also the $\Gamma$ direct bandgap being rather small (0.75~eV) by comparison to that of silicon (3.4~eV), band-to-band tunneling is strongly enhanced thus increasing dark count rates.
To reduce this further, most Ge-on-Si SPADs are designed to separate the photon absorption region (in Ge) from the multiplication region (in Si) \cite{thorburn2021ge}.
Note that band discontinuities in Ge-on-Si heterostructure (type-II, \cite[Fig.~34]{Sze2007}) do not limit the photocarriers from drifting from the absorber to the multiplication region in comparison to InGaAs/InP SPAD structures \cite{thorburn2021ge}. In Ge-on-Si SPAD, the field in Ge is reduced limiting both generation and tunneling rate in that region but still allowing for photoelectrons to drift to the multiplication region.
The multiplication region being in Si suffers only from the lower thermal generation rate of that material.
There exists many variations of separated absorption-multiplication structures (SAM, SAA, SACM, SAGCM...).
Even with optimized structures, drastically reducing the operating temperature of Ge devices\footnote{This has the unfortunate consequence of drastically reducing absorption at the bandgap edge for the infrared communication C-band (centered around 1550~nm) \cite{vines2019high}.} is unavoidable to reduce dark count rates.


Nevertheless, Ge-on-Si SPADs with 100~$\mu$m diameter where shown to provide up to 38\% of single-photon detection efficiency at a wavelength of 1310~nm, an excess bias of 5.5\% (breakdown voltage $\sim$40~V) and a temperature of 125~K \cite{vines2019high}. Smaller area Ge-on-Si SPADs (26~$\mu$m dia.) achieved a dark count rate on the order of 10$^4$ cps at similar operating bias and temperature \cite{llin2020high}.
See Thorburn~\textit{et~al.} for a broader review of performances \cite{thorburn2021ge}.

\medskip
As is the case for APDs, III-V based SPADs all have a separated absorption-multiplication structure.
The InGaAsP semiconductor family offers the possibility to grow small gap semiconductors on a larger bandgap binary member (InP) of the same family and with a well matched crystal lattice \cite[Fig.~32]{Sze2007}.
Most notable are In$_7$Ga$_5$As$_{64}$P$_{56}$/InP (quaternary compound) and In$_{53}$Ga$_{47}$As/InP (ternary compound).
Because they are direct bandgap semiconductors, considerations similar to those stated above apply to thermal and trap-assisted generation rate.
In InP, which is used for the multiplication region, holes have the largest ionization coefficient \cite{zappa1996temperature}. Therefore, the multiplication region should have its n-type side facing the InGaAs absorber region for holes to be dominant in the triggering of the avalanche in Geiger mode. 
In these heterostructures, the large band discontinuities oppose the transition of the photocarriers (holes) from the absorption to the multiplication region.
The device's conduction and valence bands must be properly engineered to effectively lower the discontinuities and mitigate their detrimental effects.

Among the first published works on InGaAsP/InP based SPADs are those of Levine~\textit{et~al.} at AT\&T Bell Laboratories \cite{levine1984near}, Zappa~\textit{et~al.} at Polytechnic University of Milan in collaboration with EG\&G Canada \cite{zappa1994nanosecond} and also McIntosh~\textit{et~al.} at MIT's Lincoln Laboratory \cite{mcintosh2002ingaasp}.
To our knowledge, the MIT's LL group was the first to report the digital readout of an array of InGaAsP/InP SPADs \cite{mclntosh2003arrays,verghese2005geiger,itzler2010design}. 
Since then, many have reported on single SPAD devices optimization \cite{signorelli2021low}. Notable works on recent SPAD arrays are those by Hamamatsu \cite{baba2018development} and MIT's LL \cite{aull2017large}.

Major challenges pave the path to SPAD array for large sensitive area devices and to integration with advanced electronic readouts as needed for imagers and PDCs.
Growth of III-V on InP is now reaching the 100~mm wafer size.


Primary sources of dark count in InGaAs/InP SPAD are tunneling through defect levels in the InP avalanche region and thermal generation in the InGaAsP absorber region \cite{donnelly2006design}. Small diameter state-of-the-art InGaAs/InP SPADs were reported with dark count rates ranging from 10$^2$~cps to 10$^4$~cps at excess voltage between 2~V and 7~V and temperatures reachable with thermoelectric cooling (225~K~-~253~K) \cite{signorelli2021low}. For similar excess voltage and temperature, InGaAs/InP SPADs have shown photon detection efficiency around 40~\% in the range of 950~nm to 1550~nm with timing jitter close to 80~ps~FWHM \cite{tosi2012fully, mpd_pdmir}. However, a major limitation to III-V SPAD detectors is the long-lived afterpulsing events affecting the maximum attainable count rate. The impact of afterpulsing can be reduced by operating the detectors in a gated mode where the SPAD is turned on synchronously with the optical signal and turned off during long period of time in order for afterpulsing events to fade away \cite{scarcella2014ingaas}.

\medskip
Although improvements in device structures and material growth contribute to continuously increasing detector performances with respect to sensitivity and dark count rate, small bandgap semiconductor infrared detectors will likely always suffer from dark count rate much higher than their silicon counterpart, even while operated at low temperatures.
Nevertheless, they can provide significant sensitivity between 900~and~1600~nm, wavelengths at which silicon cannot.
To further increase performances, one must also consider the readout techniques.
Of particular interest for large area detectors are photon-to-digital converters (a.k.a. digital SiPM) with one-to-one coupling between a SPAD and a quenching and readout circuit.
This allows an increased control on the SPAD to perform afterpulse mitigation and gating techniques that effectively reduce dark count rates.
This is discussed further in section~\ref{sec:PDC}.

\subsection{Outside HEP}\label{sec:other}

\subsubsection{Quantum Sensing}

The uncertainty when estimating the absorption/transmission of a sample --compared to a measurement using classical light-- is given by the combination of random fluctuations inherent in the optical probe beam, and by the stochastic nature of the interaction between light and matter within the sampled object. To improve the precision of such measurement is of utmost importance in Quantum Imaging. 

Recently, much attention has been paid to the field of quantum metrology and its applications in biological sciences \cite{giovanetti2011,taylor2016quantum}, and particularly to the utilization of quantum light as a resource for surpassing the classical limits of precision per unit intensity \cite{berchera2019quantum,moreau2019imaging,Brida:10}.
 Schemes for estimating the transmission of a sample generally consist in measuring the intensity attenuation of a light beam that propagates through it, which can be done using a single light beam as the source (direct measurement), or twin-beams: one arm as reference and the other as probe (differential measurement).

The best performance achievable with classical light is obtained using a beam with Poissonian statistics in the number of photons. The ultimate precision for such  direct measurement is usually called the \textit{Shot-Noise Limit} (SNL).  Through literature, twin-beams and difference-based estimators have been used for spatially resolved implementations \cite{brida2010experimental} including the realization of the first sub-shot-noise wide field microscope in 2017 \cite{samantaray2017realization}. 
The performance of this technique depends upon the spatial resolution and reaches out a factor of improvement in precision over the SNL of approximately 1.30. 
Recently, a complete theoretical and experimental study of the performance of different estimators using twin-beams was presented \cite{losero2018unbiased}, achieving a maximum improvement factor of 1.51.

State-of-the-art studies in general use conventional CCD cameras for detection, with high efficiencies $(>90 \%)$ and a readout noise usually between 2 and 5 e$^-$. These specifications impose the intensity regime in which it is possible to work: scenarios of a few photons $(\approx 50)$ per pixel cannot be explored in order to obtain an acceptable signal-to-noise ratio. 

Using sensors that enables sub-electron noise in the readout of the pixel charge, number-resolving photons in the optical and near-infrared wavelengths, would open the possibility of exploring ultra-low light intensity regimes of a few photons per pixel in transmission/absorption measurements -- which is particularly important for measuring biological samples \cite{taylor}. 

The above-mentioned applications do not require time resolution and account with the potential to produce an unprecedented impact in this field, for instance, in Quantum Microscopy \cite{samantaray2017realization, Magnoni:21}. Nevertheless, when time resolution is available, another wide universe of opportunities arises, as using coincidence, stray light reduction and background rejection becomes straightforward. 

\subsubsection{Basic Energy Science}

The photon energies of 20-500~eV offer many exciting opportunities in the basic energy sciences, but the challenge of single photon detection has hindered progress. With the Skipper-CCD, the ability to detect single photons in this unique wavelength range can offer many opportunities for new directions in this area of research. One area of interest is centered around controlling and exploiting fluctuations in quantum matter for the design of bulk materials with novel functionality. Novel x-ray methods including resonant inelastic x-ray scattering (RIXS) \cite{Ament-RMP-Ament} and x-ray photon fluctuation spectroscopy (XPFS) \cite{Shen-mrs-2021} have been developed to tackle these challenges. Both techniques are photon hungry, making single photon counting critical for extracting out the fluctuations in the system being studied. This has been demonstrated at the Linac Coherent Light Source X-ray free electron laser at SLAC \cite{Seaberg-prl-2017,Mitrano-sa-2019}, but remains in unexplored area in the energy range below 500~eV. The ability to detect single photons at lower energies will open numerous scientific applications, and could act as a catalyst to broaden further an already active area of research by enabling spontaneous fluctuation studies in new systems yet to be investigated.

\begin{figure}
    \centering
    \includegraphics[width=.5\linewidth]{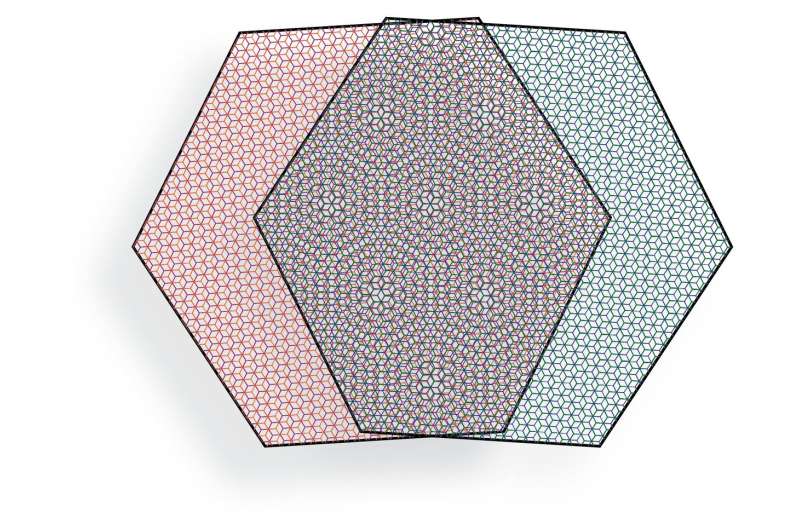}
    \caption[Alternative caption text]{Illustration of a Moiré pattern that emerges upon stacking and rotating two sheets of bilayer graphene. New physics can emerge in twisted double bilayer graphene over a small range of twist angles, and can be tuned with an electric field \cite{he-2021-natphys} (Image taken from: \hyperlink{link}{https://phys.org/news/2020-10-spontaneous-symmetry-bilayer-graphene.html}).
    }
    \label{fig:moire}
\end{figure}

For instance, one of the most interesting topics at present in condensed matter is that of twisted graphene and `twistronics' -- which could form a unique type of electronic devices \cite{Carr-prb,Cao2018}. This is based on the dramatic change of the properties of a material composed of two sheets of atomic layers by twisting each layer with respect to each other in a precise way (See Fig.~\ref{fig:moire}). Detectors which deliver single photon sensitivity at the carbon K-edge, would allow low-energy RIXS and XPFS studies of these types of systems to look at fluctuations related to the superconductivity, i.e. resistance-free current flow that can harvest energy storage, and how this phenomenon differs from the more well-known type of superconductors. Analogous to the high-temperature superconductors based on copper \cite{Keimer2015}, these studies would offer an unprecedented opportunity to investigate the many features in the phase diagram of this family of materials related to twisted graphene and could add fresh insight into an already burgeoning field.

Another area in the field of materials is that of topological matter. Here the mathematics field of topology has found an intersection with the study of materials and is shedding new insights into quantum mechanics. For instance skyrmions, particles named after the British physicist Tony Skyrme who hypothesized their existence in nuclear physics \cite{skyrme-1962-nucphys}, have been found to occur in magnetic systems and form the basis for new types of technology \cite{Fert2017}. One outstanding question is how the topological nature of these particles can affect their fluctuations? This question is critical to understanding the noise of skyrmion-based devices. With the Skipper-CCDs, single photon detection at the M-edges of 3d transition metal elements, known to have strong dichroic effects in magnetic materials, would be possible candidates to study. This could offer the potential to use two-color capabilities at XFEL facilities, such as the FERMI XUV free-electron lasers in Trieste, Italy \cite{Ferrari2016}, to study fluctuations of different elements simultaneously. This could lead to an understanding of how topology will modify ultrafast dynamics of these magnetic textures.

A third area, and one important for global alternative energy solutions, is the Nobel-prize-winning science behind Li-batteries \cite{Goodenough2010}. One of the biggest challenges in this area is realizing a high Li$^{+}$ cation conductivity in an electrolyte and across the electrode/electrolyte interface. This requires careful material engineering and screening processes. In theory, this diffusion process can be directly probed by XPFS by tuning specifically to the Li ions for direct measurement. However, an element-specific screening is not possible so far, due to the lack of a detector with single-photon sensitivity that operates at the Li K-edge (54.7 eV). Skipper-CCDs would also have impact in this broader field of battery research and technology.

Even traditional devices such as Si-based semiconductors will significantly benefit from the availability of such detectors. Charge carriers in semiconductors constitute the foundation for many key technologies at present, including computers, lasers and light emitting devices. The ability to probe their dynamics at the meV energy level, or equivalently, in the femto-to-pico second time window, is in high demand. Performing RIXS and XPFS measurements at different Si L-edges holds potential for measuring the orbital-selective carrier dynamics in these devices as well. 

\subsubsection{Applied Radiation Detection}

Photon counting detectors (CCDs) have been demonstrated as a powerful tool to measure low levels of isotopic contamination in materials as part of their use in direct dark matter detection \cite{sensei2018}. Some ideas are now being considered for implementing this detector technology for the monitoring of contamination of isotopes with low energy signals ($\sim$ 1 keV). Further developing of photon counting technologies like skipper-CCDs would benefit these applications.   

\section{Instrumentation Opportunities}\label{sec:techs}

In this section we we describe how the new technologies are addressing the needs described in the previous section, with an attempt to estimate the timescale of the developments.

\subsection{MKIDs}\label{sec:MKID}

\begin{figure}
\begin{center}
\vspace{-0.35in}
\includegraphics[width=0.7\columnwidth]{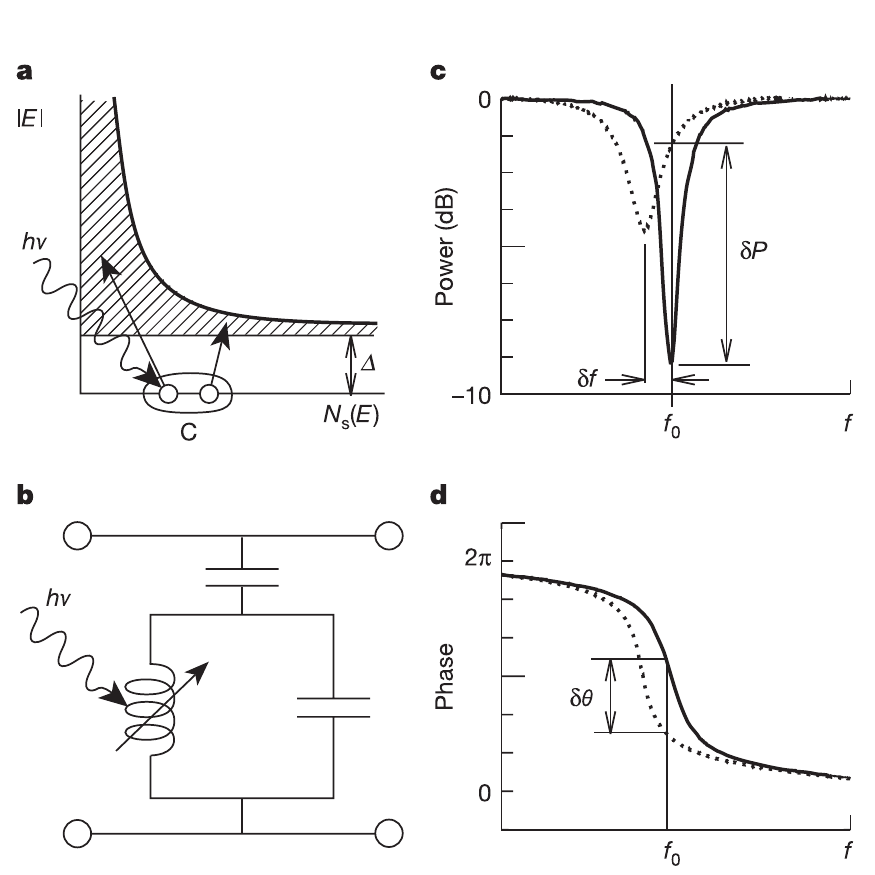}
\end{center}
\vspace{-.25in}
\caption{The basic operation of an MKID, from \cite{2003Natur.425..817D}. (a) Photons with energy $h\nu$ are absorbed in a superconducting film, producing a number of excitations, called quasiparticles.  (b) To sensitively measure these quasiparticles, the film is placed in a high frequency planar resonant circuit.  The amplitude (c) and phase (d) of a microwave excitation signal sent through the resonator.  The change in the surface impedance of the film following a photon absorption event pushes the resonance to lower frequency and changes its amplitude.  If the detector (resonator) is excited with a constant on-resonance microwave signal, the energy of the absorbed photon can be determined by measuring the degree of phase and amplitude shift.} 
\label{fig:detcartoon}
\vspace{-0.2in}
\end{figure}

MKIDs (microwave kinetic inductance detector) work on the principle that incident photons change the surface impedance of a superconductor through the kinetic inductance effect~\cite{Mattis:1958vo}.  The kinetic inductance effect occurs because energy can be stored in the supercurrent (the flow of Cooper Pairs) of a superconductor.  Reversing the direction of the supercurrent requires extracting the kinetic energy stored in it, which yields an extra inductance term in addition to the familiar geometric inductance.  The magnitude of the change in surface impedance depends on the number
of Cooper Pairs broken by incident photons, and hence is proportional to the amount of energy deposited in the superconductor. This change can be accurately measured by placing a superconducting inductor in a lithographed resonator, as shown in Figure~\ref{fig:detcartoon}.  A microwave probe signal is tuned to the resonant frequency of the resonator, and any photons which are absorbed in the inductor will imprint their signature as changes in phase and amplitude of this probe signal.  Since the quality factor $Q$ of the resonators is high and their microwave transmission off resonance is nearly perfect, multiplexing can be accomplished by tuning each pixel to a different resonant frequency with lithography during device fabrication.  A comb of probe signals can be sent into the device, and room temperature electronics can recover the changes in amplitude and phase without significant cross talk~\cite{2012RScI...83d4702M}.

After a decade of development UCSB is currently producing high quality PtSi and Hf MKID arrays.  These are the best optical to infrared (OIR) MKID arrays ever produced with up to 20,440 pixels, ${\sim}$80\% of the pixels functional, R=E/$\Delta$E$\sim$9.5 at 980 nm, and a quantum efficiency of ${\sim}$35\%. These state-of-the-art MKID arrays are discussed in detail in Optics Express~\cite{Szypryt:2017cb} and in the instrument paper of MEC, a 20 kpix MKID camera permanently deployed at the Subaru Telescope on Maunakea~\cite{2020PASP..132l5005W}.

\begin{figure}
	\vspace{-0.2in}
  \centering
  \includegraphics[width=0.7\columnwidth]{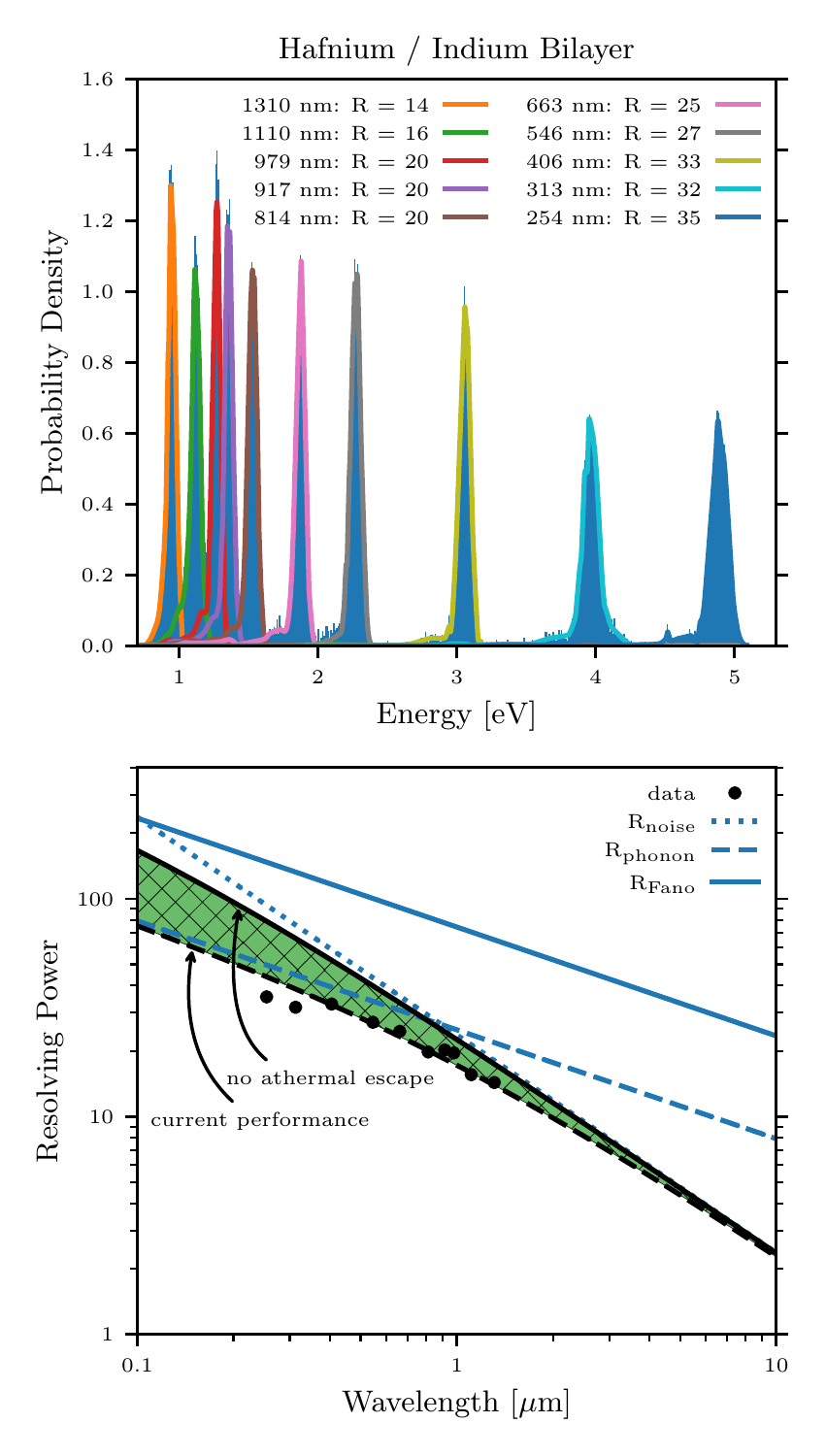}
\vspace{-0.3in}
  \caption{Top: The combined spectra for the Hf/In bilayer device at seven laser frequencies. Bottom: The noise decomposition for this device. The filled in green area represents the R achievable by reducing the phonon loss but keeping the same noise spectrum.}
  \label{fig:bilayer}
  \vspace{-0.2in}
\end{figure}

Recent breakthroughs in the understanding of MKID noise, in conjunction with a quantum-limited parametric amplifier~\cite{2019ApPhL.115d2601Z}, have dramatically improved MKID spectral resolution, as shown in Figure~\ref{fig:bilayer}. We expect this to improve further in the next several years as we optimize designs that eliminate the primary noise source, athermal phonon escape, and approach the Fano limit.

\subsection{SNSPDs}\label{sec:SNSPD}

A Superconducting Nanowire Single Photon Detector (SNSPD) is a superconducting film patterned into a wire with nanometer scale dimensions (although recently devices with micrometer-scale widths have been shown to be single-photon sensitive \cite{Korneeva2018}). A typical SNSPD is made from a film roughly 5~nm thick and patterned into a meandering wire roughly 100~nm. Arrays of thousands of SNSPDs have been fabricated with 1 mm$^2$ total active area \cite{Wollman:2019}. A bias current, $I_B$ is applied to the wire that is slightly below the nanowire’s critical current, $I_C$ --the current that will drive the nanowire from the superconducting to resistive state. Due to the presence of this bias current, the energy of a single photon is sufficient to drive the superconducting nanowire into its resistive state. Depending on the readout configuration, the bias current is either shunted to a bias resistor, or drops due to the high resistance of the wire. With the reduced bias current, the nanowire returns to its superconducting states and is ready to detect another photon. These devices offer a wide-wavelength detection range and can be tailored to specific optical detection needs. Their lack of a shielding requirement, off-the-shelf microwave amplifier readout, and high-temperature (T$>$1K, i.e. relative to other superconducting detector technologies) operation make them straightforward to implement for a wide variety of applications.

SNSPDs have been reported with single photon sensitivity for wavelengths out to several microns, timing jitter as low as a few~ps \cite{Korzh:2018}, dark count rates (DCR) down to $6\times10^{-6}$ Hz \cite{Chiles:2021}, and detection efficiency (DE) of 0.98 \cite{Reddy:2020}. They have also been  have been shown to function in magnetic fields of up to 6T \cite{Lawrie:2021}.

These properties are determined by several factors, both intrinsic to the detector (e.g., material choice and device design) and from external operating settings (e.g., bias current, and operating temperature). Often these parameters present a trade-off when optimizing the detector performance. Detection efficiency is primarily influenced by the optical coupling scheme (if a photon is absorbed, it can often be detected), but benefits from a lower energy gap for a given operating temperature and limited constrictions---from either material grain size or detector geometry---which can limit the applied bias current. The dark count rate is optimized by limiting sources of fluctuations---either thermal or bias current---and benefits from a larger critical temperature ($T_C$), energy gap, and critical current density -- conditions opposed to those benefiting detection efficiency. Thus the dark-count rate is often optimized through operating parameters such as a lower operating temperature and bias current. Timing jitter performance seems to benefit from materials with a high $T_C$ (related to larger $I_C$ which allows a larger $I_B$) and normal resistance---a larger normal resistance results in larger signal to noise which enables reduced jitter---and a low inductance which allows for a faster rise time. Jitter performance also benefits from the use of low noise readout electronics, such as low noise cryogenic amplifiers.

SNSPDs have been made from a wide range of materials with NbN and WSi two of the most common. NbN films were used in the very first nanowires and have been widely used ever since. NbN provides a high $T_C$ - around 15~K for thick films, but reduced to 7-8~K for the very thin films used for nanowires. WSi films can have $T_C$ as high as 5~K and offer tunable $T_C$, resistivity, and inductance based upon Si content. A key difference between the two films is crystalline structure. NbN films offer a polycrystalline film and often require a high temperature growth process to yield larger grain sizes, and the lower resistances desired for good jitter performance. WSi films are amorphous, and with a smaller energy gap, have yielded some of the highest detection efficiency results. However, their corresponding higher resistance and inductance generally results in a higher jitter and longer reset time than NbN. Other materials (e.g. NbSi or MoSi) are being investigated for their amorphous structure and may offer lower inductance than WSi.  Some efforts have been made to realize SNSPDs with still higher $T_C$ by using MgB$_2$\cite{Cherednichenko_2021} ($T_C > 30 K$ in thin films) and YBCO\cite{ARPAIA201516} ($T_C > 70K$ in thin films).

SNSPDs are widely implemented across a broad spectrum of Quantum Information Science and Technology research including quantum communication and quantum sensing. SNSPDs also have direct applications in HEP \cite{Polakovic2020}. They are used as detectors for foundational experiments utilizing quantum teleportation and communication protocols and recently, they are being explored as novel detectors for dark matter searches\cite{Chiles:2021,Hochberg2019}. 

The scalability of row/column readout scheme, such as those employed by the current kilopixel arrays \cite{wollman2019kilopixel}, is limited by a corresponding reduction in maximum occupancy, signal amplitude, and signal-to-noise \cite{allman2015near}. As the number of pixels scale up toward the megapixel mark, cryogenic readout becomes essential to limit the complexity of the system. 

As part of one of DOE's recent Microelectronics Co-Design Research programs, Fermilab is currently leading a collaborative effort to enable the scaling of large SNSPD arrays, coupled to cryogenic custom readout, both superconducting and cryoCMOS.
The resulting custom ASICs and superconducting readout will handle fast multi-channel timing and sensor multiplexing directly at cryogenic temperatures, which should improve the signal to noise of the system and preserve the timing performance of the sensor. Other benefits include the increase in maximum rates through the use of active quenching techniques \cite{ravindran2020active}, and the potential to achieve multi-photon resolution, for example via time-over-threshold measurement, since the risetime has been shown to be inversely proportional to the square root of the simultaneous detection events \cite{cahall2017multi}. 
Through such developments, as well as recent work demonstrating SNSPDs with wide wire geometries that are amenable to standard optical lithography, very large SNSPD pixels and large arrays of SNSPDs will continue to open new applications for HEP.

\subsection{Germanium Detectors}\label{sec:gedet}

\begin{figure}
\begin{center}
\includegraphics[width=0.7\hsize]{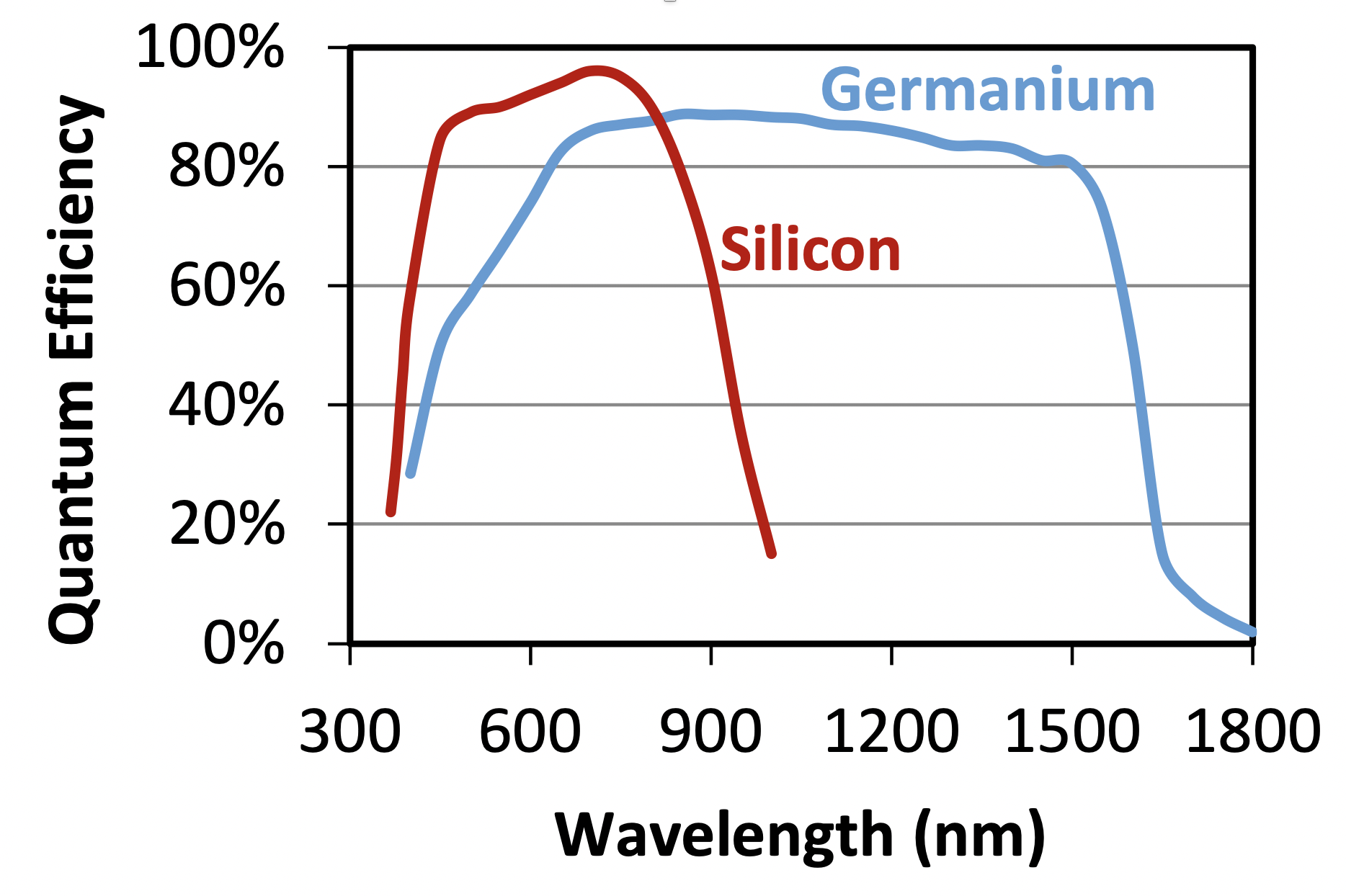}
\caption{Representative quantum efficiency curves for silicon and germanium detectors~\cite{Nakano2012,Posthuma2003}.}
\label{fig:Ge00}
\end{center}
\end{figure}

    Silicon CCDs are commonly utilized for scientific imaging applications in the visible and near infrared. These devices offer numerous advantages described previously, while the skipper CCD~\cite{tiffenberg2017single} adds to these capabilities by enabling multiple samples during readout to reduce read noise to negligible levels~\cite{AguilarPRL2017,sensei2018}. CCDs built on bulk germanium offer all of the advantages of silicon CCDs while covering an even broader spectral range (Fig.~\ref{fig:Ge00}). Furthermore, while this technology is still under development, additional features such as skipper readouts or orthogonal transfer (to correct for distortions due to atmospheric turbulence) could be added to future generations of these devices.

A front-illuminated germanium CCD was first demonstrated in the 1970s~\cite{Schroder1974}, and later further explored in the 1990s~\cite{GeCCD}, but these devices suffered from a variety of limitations such as high dark-current and poor charge-transfer efficiency (CTE). However, recent improvements to germanium material quality and advances in surface passivation now enable germanium CCDs with performance attributes that match silicon CCDs. High-quality bulk germanium wafers are now commercially available in 150- and 200-mm diameters, enabling device fabrication in the same facilities used to process silicon detectors. Furthermore, gate dielectrics with low surface state density can now be grown on germanium, generally through growth of thin layers of GeO$_2$ via thermal oxidation and subsequent encapsulation and protection during processing to avoid etching or decomposition of this critical film. This enables high charge-transfer efficiency and low surface dark current, both critical for operation of scientific CCDs.

\begin{figure}
\begin{center}
\includegraphics[width=.9\hsize]{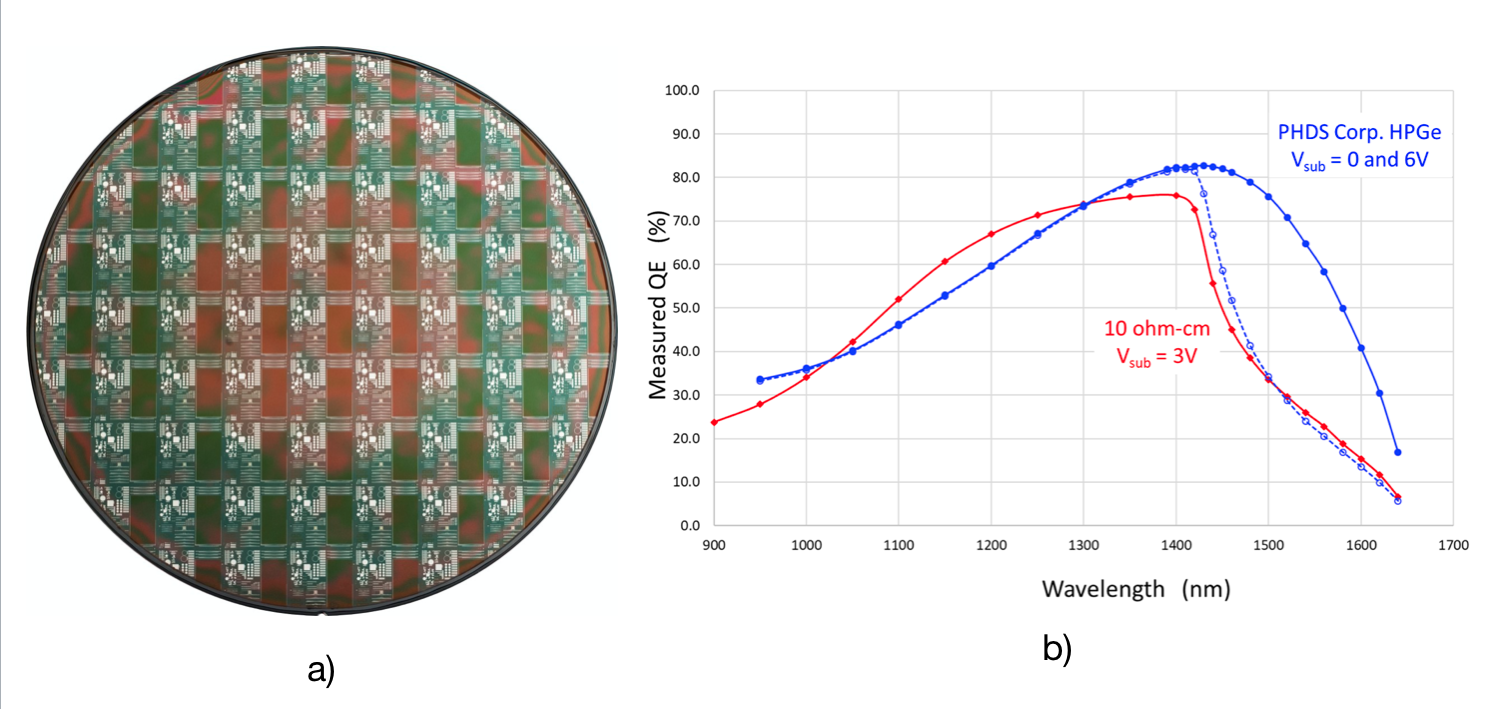}
\caption{a) Image of CCDs fabricated at MITLL on 200-mm diameter germanium wafers \cite{Leitz2020} .  b) Measured quantum efficiency on
1~cm$^2$ photodiodes fabricated on 150-mm diameter germanium wafers at Lawrence Berkeley National Laboratory.  The depletion
depths are about 10~$\mu$m for the 10 ohm-cm Ge, and 650~$\mu$m (fully depleted) for the HPGe wafer fabricated from a PHDS
Corp. crystal and operated at a substrate bias voltage of 6V.  The operating temperature was 100K.}
\label{fig:Ge01}
\end{center}
\end{figure}

Germanium CCDs have been under development at both  MIT Lincoln Laboratory (MITLL) and Lawrence Berkeley National Laboratory (LBNL).  MITLL has realized kpixel-class front- and back-illuminated 
devices~\cite{Leitz2018,Leitz2019,Leitz2020} fabricated on 200-mm wafers 
(Fig.~\ref{fig:Ge01}~a). The current effort at MITLL is focused on incorporating design and process improvements aimed at demonstrating low read noise, charge-transfer efficiency suitable for scientific detectors, and improved yield, with the goal of eventually matching the capabilities of silicon CCDs on these key metrics.  LBNL has worked with PHDS Corp., a commercial company that produces gamma-ray detector systems, on the development of 150-mm diameter high-purity Ge (HPGe) wafers.  PHDS Corp. has 
the capability to grow large diameter, high-purity crystals, and through a Small-Business Innovation Research grant 150-mm 
wafers were produced at Umicore from the PHDS Corp. cystals.  Figure~\ref{fig:Ge01}~b) compares the quantum efficiency of 1~cm$^2$ photodiodes fabricated on 150-mm diameter Ge wafers at LBNL.  The 10 ohm-cm devices have a depletion depth of about 10~$\mu$m, while the 650~$\mu$m-thick HPGe diode is fully depleted at 6V substrate bias voltage.  The near-infrared 
quantum efficiency extends to longer wavelengths for the HPGe device as expected.

Finally, germanium can also be utilized in hybrid active-pixel sensors. In this context, germanium offers an important potential advantage over compound semiconductors: transfer gates and/or amplifiers can be incorporated onto the detector tier. This enables separation of charge collection and readout, as is the case in hybrid silicon scientific active-pixel sensors, which should enable read noise of a few electrons in a radiation-tolerant device with 100\% fill factor.

\subsection{Skipper-CCDs}\label{sec:skipper}

\begin{figure}
\begin{center}
\includegraphics[width=1\hsize]{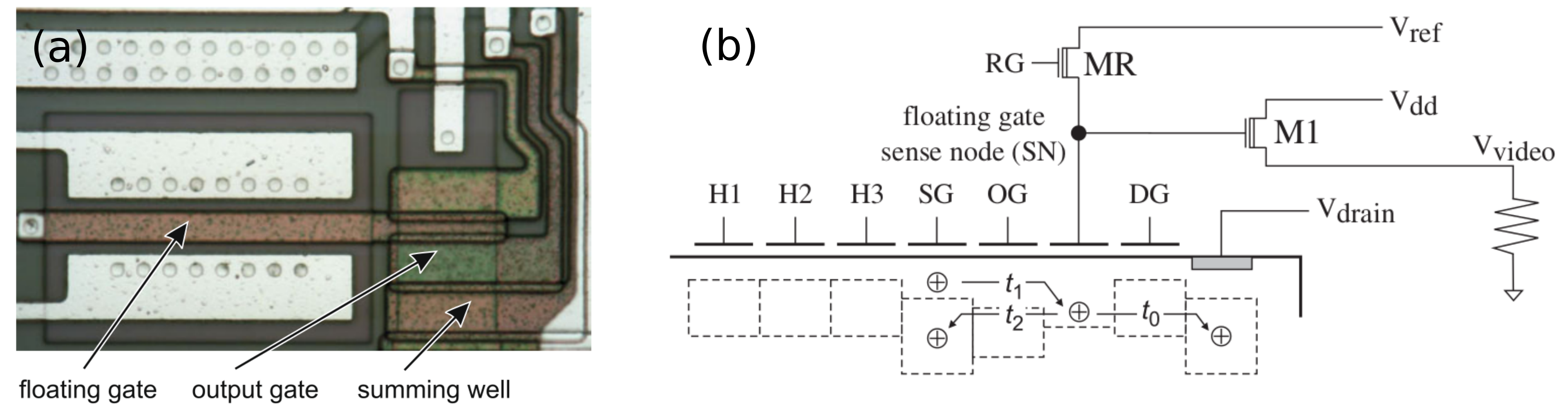}
\end{center}
\caption{(a) Microscope view of the non-destructive readout output stage of the Skipper CCD. (b) Typical charge transfer operation of the Skipper CCD in the output stage. Figures were taken from \cite{skipper_2012}.}
\label{fig:floating gate skipper}
\end{figure}

\begin{figure}
\begin{center}
\includegraphics[width=0.7\hsize]{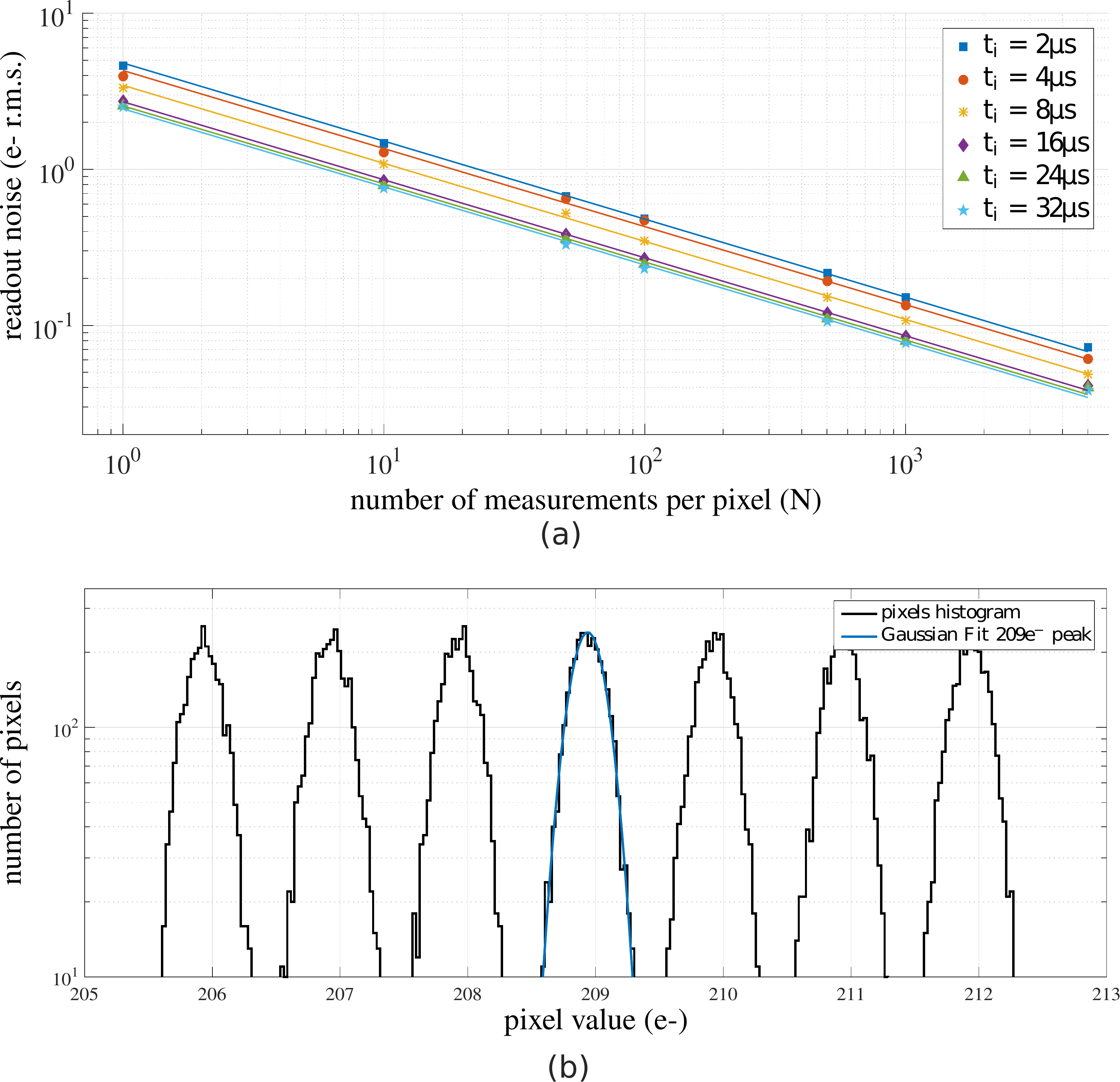}
\end{center}
\caption{(a) noise reduction as a function of the number of samples per pixel. The different colors are measurements for different integration time per sample. The solid line is the best fit of the theoretical model to the points. (b) Pixel histogram of an image taken with deep sub-electron noise. Figures were taken from \cite{lta_2021}.}
\label{fig:noise reduction skipper}
\end{figure}

Skipper-CCDs have an output readout stage that allows multiple non-destructive sampling of the charge packet in each pixel of the array thanks to its floating gate output sense node (shown in Figure \ref{fig:floating gate skipper}(a)). The charge of each pixel is moved to the sense node (SN) where it is measured by the output transistor. Figure \ref{fig:floating gate skipper}(b) shows its operation. After the charge is measured, it is moved back to the summing-well gate (SG) while the voltage of the SN is reset using the reset transistor (MR). At this time the SN is ready to take a new sample of the same charge packet. This process can be repeated many times. All charge measurements have an independent readout noise contribution, and therefore its final pixel value is calculated as the average of the available samples \cite{skipper_2012, Tiffenberg2017}. Figure \ref{fig:noise reduction skipper}(a) shows the reduction of the standard deviation of the noise error in the pixels as a function of the number of samples per pixel. The different colors correspond to different time periods (t$_i$) spent to read each sample. The reduction factor is the square root of the number of samples.  Figure \ref{fig:noise reduction skipper}(b) shows the histogram of pixels with a charge from 206 to 212 carriers with deep sub-electron noise where the number of carriers can be discretized with no ambiguity. The charge counting capability is similar in the entire dynamic range of the sensor from zero to the maximum full well capacity of the pixels.

The Skipper CCD fabricated on high resistivity silicon \cite{Holland_2003} has also demonstrated an extremely low production of dark counts. Recent measurements in \citep{PhysRevLett.122.161801} demonstrated production rates close to $10^{-4}$ e$^-$ per pixel per day with a pixel size of 15 $\mu$m by 15 $\mu$m and a thickness of 675 $\mu$m.

Since the single counting capability was proven \citep{tiffenberg2017single}, it has motivated to build a new generation of Dark Matter \cite{oscura_2020, PhysRevLett.122.161801} and neutrino experiments \cite{Nasteva_2021, fernandez2021LightMediators} that will be at the forefront of exploring physics beyond the Standard Model. The \textit{Sub-Electron Noise Skipper-CCD Experimental Instrument} (SENSEI), has produced  world-leading constraints on low-mass dark matter searches \citep{PhysRevLett.122.161801}. Skipper-CCDs have also been identified as a powerful tool for optical applications such as quantum information imaging giving access to entangled measurements in momentum and spatial variables for single photons, and astronomical applications where it provides an attractive approach to reduce readout noise while keeping the beneficial characteristics of conventional CCDs.

Since the noise reduction in the sensor involves spending extra time on each pixel, the readout speed is currently a limiting factor for some applications. There are several ongoing efforts to improve this aspect including parallelization of the array readout through multiple single-amplifier output stages; multiple readout amplifiers on a single readout stage, frame-shifting architectures (where readout and exposure can be done at the same time), or dynamic skipper readout, where photon counting is only targeted over a subset of the detector pixels \cite{smart_skipper_2021}.


\subsection{Skipper-CMOS}\label{sec:cmos}


The extremely low readout noise of Skipper-CCDs allows the detection of single photons in the optical and near-infrared range. Unlike other silicon detectors with an avalanche gain, with Skipper-CCDs it is possible to count the exact number of electrons per pixel and therefore the number of photons that interacted on each pixel, being only limited by the Fano noise \cite{Janesick1988} \cite{rodrigues2020absolute}. Skipper-CCDs has been identified as a powerful tool for quantum information science, giving access to entangled measurements in momentum and spatial variables for single photons, with the initial demonstration recently completed at Fermilab.

Skipper-CCDs are fabricated in a dedicated facility using a customized process for scientific CCDs \citep{ccdProcess}. This process is required to produce the overlapping of the gates structures needed to  achieve high charge transfer efficiency between pixels. Due to the very low demand of scientific CCDs, compared to commercial CMOS imagers, the number of facilities dedicated to scientific CCDs has been reduced to only a few in the world today, and this number is expected to continue dropping \citep{dawson2019maintaining}. On the other hand, imagers fabricated in CMOS process have dominated the market of high-demand consumer cameras, and therefore several fabrication facilities with many processing options are available. Moreover, previous works have successfully implemented CCDs in different single-poly CMOS fabrication technologies achieving high charge transfer efficiencies \cite{boulenc2017high}\cite{crooks2013kirana}\cite{marcelot2014study}\cite{fife2009design}, and the possibility to implement Skipper-CCDs in CMOS have been recently analyzed in \cite{stefanov2020simulations}. 

The scaling of CMOS technology of at least 100\,nm has allowed the implementation of pixels with a very low capacity, and therefore, high sensitivity and low noise (1-2\,$\rm e^-$) at room temperature and high frame rates (50-100\,fps) \cite{ma20154mp}\cite{fowler20105}. There are two new developments on CMOS technologies that have achieved noises low enough to be considered photon-resolving technologies. On  2021 Hamamatsu announced their first commercial camera using what they  call “quantitative CMOS” (qCMOS). This sensor, with a proprietary processing, is able to get  a readout noise of 0.27\,$\rm e^-$ at 5\,fps, and  0.43\,$\rm e^-$ at 120\,fps, with a 9.6\,Mpix frame and a pixel size of 4.6\,$\mu m$\cite{hamamatsu_technote}. The camera electronics has an active role, by calibrating each pixel in real time in order to correct for sensitivity non-uniformity's and keeping a very stable temperature\cite{hamamatsu_whitepaper}.  The second device is a Quanta Image sensor(QIS), by Gigajot, a spin-off from Dartmouth college, where the QIS (based on binary pixels called “jots”) was developed\cite{Quanta_Fossum}\cite{QIS_whitepaper}. Their sensors achieve 0.19\,$\rm e^-$ at 30\,fps with 16\,Mpix frames and a pixel size of 1.1\,$\mu m$. Their process consisted on suppressing temporal noise from the pixel source followers and also increasing considerably the conversion  gain\cite{Quanta_camera}. 

\begin{figure}
\begin{center}
\includegraphics[width=0.7\hsize]{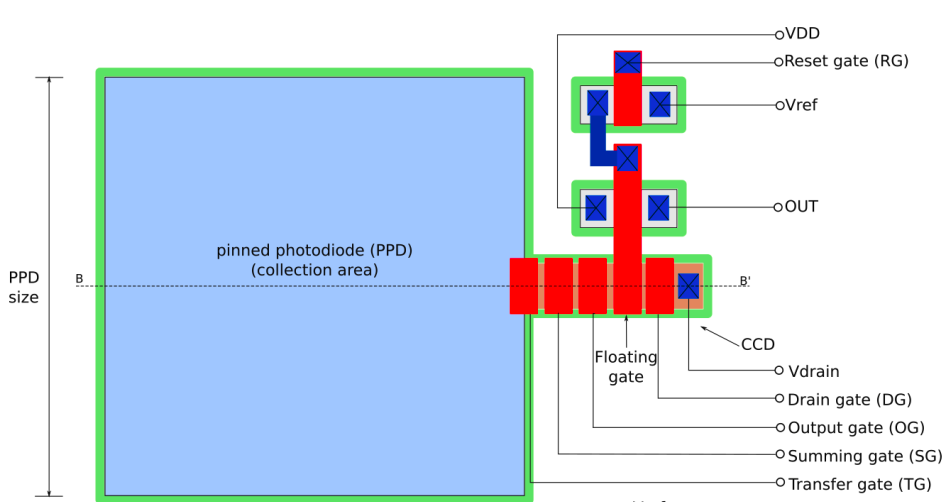}
\end{center}
\caption{Pixel concept of the Skipper-CMOS imager.}
\label{fig:skipperCMOS}
\end{figure}

The aim of the Skipper-CMOS effort is to demonstrate the non-destructive charge readout of Skipper-CCD technology in CMOS technology and benefits from the mainstream commercial integrated circuits developments. This will bring the possibility to achieve an imager composed of pixels with extremely low readout noise, that allows single-photon detection and precise counting in a wide dynamic range. Also, this effort address directly the current challenge in the fabrication of scientific Skipper-CCDs. It will have the additional advantage of allowing in-chip integration of a video processing stage (on-chip ADC), with the potential of converting the Skipper-CCD into a fully digital device.  Figure \ref{fig:skipperCMOS} shows the pixel concept of a CMOS imager with non-destructive charge readout like Skipper-CCD detectors. The collection area is performed by an pinned-photodiode (PPD), which provides low dark-current and large collection area with all the optical characteristics of the CMOS process \cite{fossum2014review}. In a similar way to 4T CMOS imagers, the charge is transfer from the PPD to the sensing stage through a gate, called transfer gate in figure \ref{fig:skipperCMOS}. Unlike CMOS imagers, the floating diffusion in the sensing stage has been replaced by the output stage of an Skipper-CCD. In similar fashion to Skipper-CCD, the output stage is composed by a buried channel CCD with four gates, where one is the floating gate (FG) used for sensing of the charge packet non-destructively. The MOSFET that shares the gate with the floating gate, is in a source follower configuration for charge to voltage conversion, and the other MOSFET is used for resetting the floating gate. The operation of the output stage, is similar to the readout of Skipper-CCDs \cite{skipper_2012}. The output stage can be covered by metal layers to shield it from the photon source. As was previously mention, with CMOS process it is possible to achieve a low readout noise, and therefore few skipper samples would be required to reach the sub-electron readout noise regime. Moreover, the high parallelization of the imager can allow image acquisition at high speeds. This development can provide the necessary detectors for future astronomy, quantum imaging and basic energy science that require increase the readout speed of the actual available Skipper-CCDs, and also could be applied in dark-matter and neutrino experiments.

\subsection{Single Photon Avalanche Diode Array Based Detectors} \label{sec:SPAD-PDC}

The single photon avalanche diode (SPAD) is an avalanche photodiode biased above its breakdown in a metastable state awaiting an avalanche to be triggered by a photoelectron or a thermally excited carrier.
This operation is often termed as Geiger-mode in reference to the metastable operation of Geiger counters.
Once triggered, a resistor or an electronic circuit quenches the current flowing through the photodiode, thus reducing the photodiode bias below the breakdown voltage. 
A SPAD is a non-linear photodetector by nature because its response is identical whether it was triggered by one or many photons. 
The SPAD’s response is Boolean: it is either waiting for a carrier in the metastable state above breakdown or it is triggered with a current flowing until quenched.

The development of SPADs started in the mid 70s \cite{1975-Cova_AQC-invention} including discrete quenching electronics with continuous improvement of performance through device understanding and better electronics \cite{TISA2007113,1993-Dautet-McIntyre_Photon-Counting,lacaita1995performance}.
In the 90s, many groups moved to integrate many SPADs into a single large-area photon counting device leading to the first SiPMs, with resistively quenched SPADs in a large parallel array in the early 2000s \cite{2000-Saveliev_first-SiPM,1997-Zappa_SPAD, 1997-Kindt_SPAD-array,1998-Kindt_SPAD-CMOS, 1999-Kindt_thesis,2003-Rochas_thesis, 2003-Rochas_SPAD-CMOS,Rochas2DSiPM,2005-Niclass-Rochas-Charbon_CMOS-SPAD-camera-32x32-800nm,2015-Charbon_dSiPM-review}.

\subsubsection{Enhancing VUV sensitivity\label{sec:SPAD_VUV}}
Silicon SPADs and SiPMs have excellent PDE in the visible spectrum.
However, sensitivity falls dramatically for wavelengths below 350~nm.
This is due first by the commonly used passivation materials which absorb strongly at these wavelengths.
But foremost is the very short penetration length of those photons (few nanometers) causing the photogenerated carriers to be trapped by surface fields and recombined through surface defects.
As mentioned before, liquid argon detectors often rely on wavelength shifting materials to circumvent this problem.
Through much internal electric field and passivation layer optimization, Hamamatsu\footnote{\url{https://www.hamamatsu.com/us/en/product/optical-sensors/mppc.html}} and FBK\footnote{\url{https://sd.fbk.eu/en/projects/detail/sipm-evolution/}} now offer SiPMs reaching 10\% to 20\% PDE at the liquid xenon scintillation wavelength (175~nm).
These are contenders for nEXO and DARKSIDE experiments \cite{Nakarmi_2020,jamil2018vuv,gallina2019characterization,Aalseth2021_DARKSIDE}.

Yet in the 1990s, Jet Propulsion Laboratory showed that CCD sensitivity can be increased closed to the reflection-limited quantum efficiency of silicon down \cite{Hoenk1992_UVCCD_Delta}.
This was done by blocking the surface fields and traps through the epitaxial growth of a strongly doped very thin silicon layer (delta-doping).
Quantum efficiency exceeding 50\% were demonstrated in CCDs down to 125~nm wavelength \cite{nikzad2012delta}.
The method was demonstrated efficient on backside illuminated SPAD based detectors by Schuette in~2011\cite{schuette2011mbe}.
Other methods to address the surface fields and traps issues were also demonstrated \cite{nanver2014robust}.
Work is being done at Caltech (D.~Hitlin) to enhance SiPMs for the detection of the fast scintillation component of BaF$_2$ \cite{hitlin2021progress}.
An extensive study of the delta-doping approach to enhance VUV sensitivity in frontside illuminated SPAD based detectors was done by Vachon \cite{Vachon20221-MScA}.

Enhancing PDE above 30\% would bring drastic performance improvements to noble liquid TPC based experiments.
Notably, it would allow direct detection of liquid argon scintillation light. 

\subsubsection{Near Infrared Enhanced Silicon Photomultipliers\label{sec:SPAD_NIR}}
The single-photon detection nature of SPAD makes it an ideal candidate for applications requiring high detection sensitivity and low temporal jitter such as LIDAR, biophotonics and quantum key distribution.
SPADs made of silicon can in theory absorb photons up to $\sim$1110~nm (E$_g =$ 1.11~eV) but practically is limited to wavelength below 900~nm. 
Nonetheless silicon SPADs are being used in the infrared wavelength range \cite{takai2016single, bruschini2019single,hadfield2009single,podmore2021qkd}.
However, the typical p$^+$n structure of SPADs is not suitable for near-infrared light. 
Indeed, the pn junction is usually found at depths of 100~nm to 1~µm while NIR photons are mainly absorbed beyond 10~µm.
Nevertheless, the exceptional qualities of silicon and its compatibility with CMOS technologies have justified manufacturers to adapt them enhancing their NIR sensitivity.

The most frequent approach is to use an n$^+$p junction with a thick p-type epitaxial layer in order to favor the collection of deep photogenerated electrons.
Increasing the detection volume increases the SPAD photon detection efficiency but at the expense of an increase in dark count rate and temporal jitter. 
It also impacts manufacturing techniques and, for example, calls for higher aspect ratio trenches to isolate SPADs when put together in an array. 
Collecting NIR photons favors backside illuminated structures where photons are incident on the thick p-type layer side (or substrate) of the device.
In this case, a structure with separate absorption and multiplication regions is generally used \cite{paternoster2019silicon, al2016backside, lee2017back}.
A thick depletion region benefits the NIR-enhanced SiPMs \cite{gulinatti2012new}. It also reduces capacitance and thus decreases correlated-type events such as afterpulsing and optical crosstalk \cite{acerbi2018silicon}. 

LIDAR systems in the automotive market has prompted several companies to produce analog NIR-enhanced SiPMs. They reach PDEs on the order of 10~to 20\%  for wavelengths between 800~and 900~nm \cite{broadcom_nirenhancedsipm, onsemi_nirenhancedsipm, hamamatsu_nirenhancedsipm, mazzillo2012electro, beer2018spad, organtini2020industrial}. Single SPAD modules with a very large diameter and thickness reached PDEs up to 40\% at 905~nm and 55\% at 850~nm \cite{excelitas_nirenhancedspad}. Such high PDE is not yet achieved in SiPMs.

\subsubsection{Photon-to-Digital Converters\label{sec:PDC}}

Single photon avalanche diodes (SPAD) are the building block of both analog SiPM and photon-to-digital converters (PDC).
An analog SiPM sums the charges of individual SPADs, passively quenched by a resistor and all connected in parallel to the output node.
In a PDC, each SPAD is coupled to its own electronic quenching circuit.\footnote{Note that digitally quenched SPADs were used before the advent of SiPMs \cite{1975-Cova_AQC-invention}.}
This one-to-one coupling provides control on individual SPADs and signals each detected avalanche as a digital signal to a signal processing unit within the PDC.
Hence, PDCs provide a direct photon to digital conversion considering that intrinsically a SPAD is a Boolean detector by design.
On the contrary, the summed current signal of an analog SiPM requires a low noise high power preamplifier, current amplifier or transimpedance amplifier followed by a shaping amplifier and analog-to-digital converter to obtain a digital information.
All this to get the Boolean information that was readily available at the sensor level.
Moreover, as the charge of each avalanche varies from SPAD to SPAD and from event to event, current fluctuations blur the signal for the same amount of photons detected and limit single photon resolution to the low count range.

Analog SiPMs have large output capacitance coming from all SPADs connected in parallel.
This needs to be considered in details while designing the front-end electronics, in particular with respect to power consumption.
In large area systems, performance compromises must be made between low power and fast time response.
This is not the case for PDCs in which the SPAD quenching can be made with very low power consumption (tens of µW).
Their output being digital, PDCs with various wavelength sensitivity (or other properties) could be assembled more easily in a system.
The direct conversion also eliminates signal variations due to SPAD-to-SPAD and event-to-event fluctuations.
A PDC therefore offers single photon resolution up to its full dynamic range (i.e.\ number of SPADs in the array).
The individual control on each SPAD allows to disable noisy cells either due to fabrication or degradation.
In a PDC one can therefore mitigate SPAD degradation, be it due to either normal ware or radiation damage, and maintain the noise floor to its minimum at the cost of marginal dynamic range loss (number of disabled SPADs).
Using an active circuit to control and readout the SPADs allows to configure and optimize the hold-off delay and recharge time according to specific applications.
In particular by properly tuning the hold-off delay, the impact of afterpulsing events can be very efficiently reduced\cite{vachon2020measuring}, a relevant feature in cryogenic detectors.
Lastly, depending on the application's needs, complex digital signal processing can be embedded in the PDC.
This can include dark count rate filters, time-to-digital converters (TDC), data compression, etc.

Digital SiPMs were first reported in 1998 \cite{1998-Aull_SPAD-3D} by the MIT Lincoln Lab and many contributions followed \cite{1998-Aull_SPAD-3D, Aull2016}.
A major step came with microelectronics integration to fabricate both the SPAD and readout quenching circuit in a single commercial process \cite{1997-Zappa_SPAD, 1997-Kindt_SPAD-array, 1998-Kindt_SPAD-CMOS, 1999-Kindt_thesis,  2003-Rochas_thesis, 2003-Rochas_SPAD-CMOS}. 
These innovations led to the first multi-pixel digitally read SPAD arrays \cite{Rochas2DSiPM, 2005-Niclass-Rochas-Charbon_CMOS-SPAD-camera-32x32-800nm}. 
Such 2D digital SPAD arrays have progressed tremendously \cite[see][and references therein]{2015-Charbon_dSiPM-review}.
When implemented in two dimensional integration (i.e.\ the SPAD along side the CMOS functions on the same chip), a compromise must be made between the area devoted to the SPADs and that devoted to the CMOS functions.
An efficient way to overcome this is to stack the SPAD array onto the CMOS readout in what is called 3D integration.
Many large players of the industry have followed this path for imagers \cite[e.g.,][]{Sony3DCMOScamera}.
3D integration allows for larger sensitive area, and more uniform connections to the SPAD in order to reach ultimate performances in timing resolution and/or power dissipation.
A review and at length discussion on the subject was published in 2021 by the authors \cite{Pratte2021_Sensors}. 
We are designing PDC based systems for various applications including noble liquid detectors (in particular LXe for nEXO), scintillation neutron detectors, positron emission tomography (PET), and QKD.
Besides QKD in the NIR, these applications use the high sensitivity of silicon in the visible spectrum and include enhancing work to enhance their VUV sensitivity.

\subsection{Photon Counting with TES} 
\label{sec:TES}
\begin{figure}[!htb]
	\centering
	\includegraphics[height=2.5in, keepaspectratio]{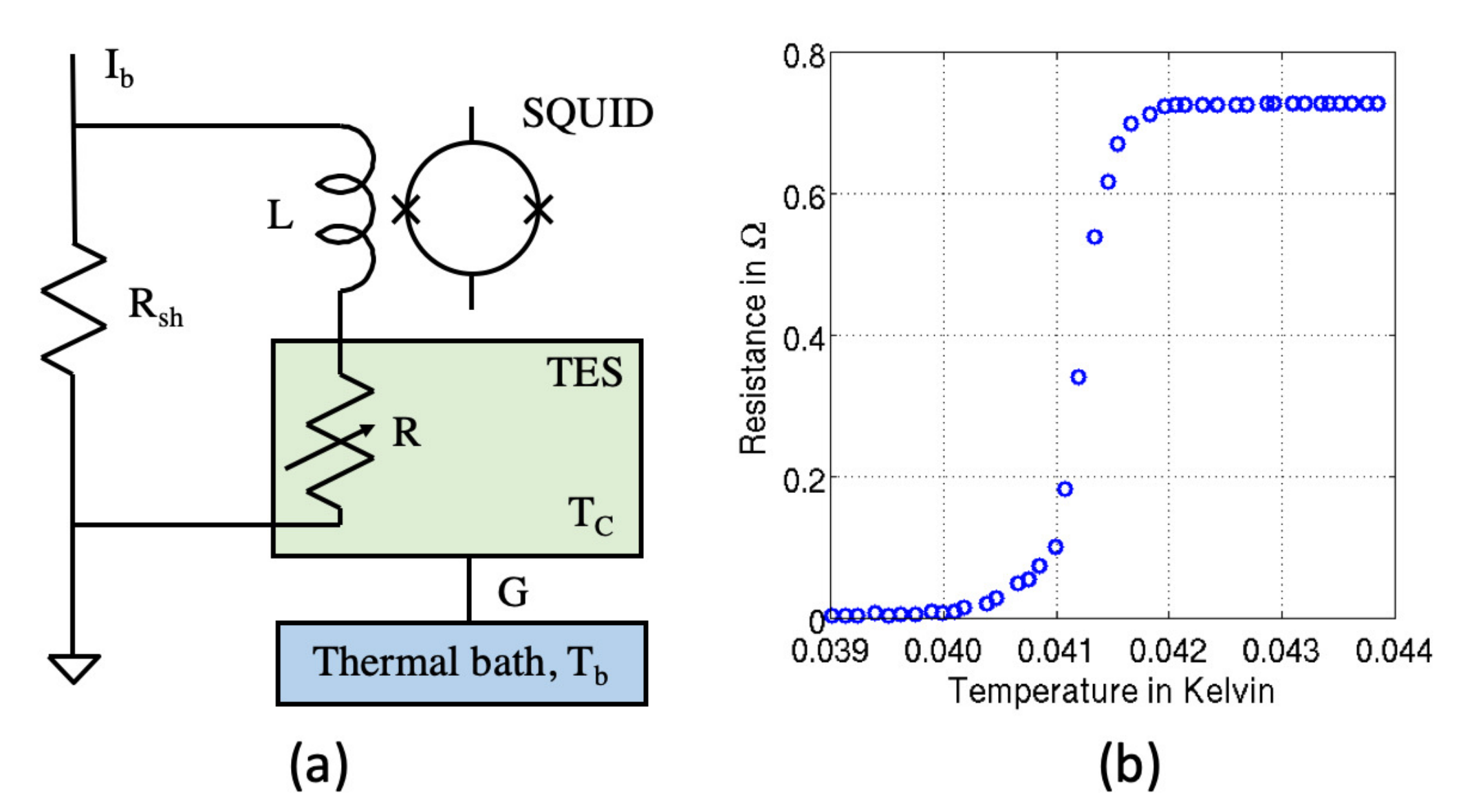}
	\caption{(a) Schematic of operation of a TES detector with a resistance $R$. The TES is in series with an inductor $L$ coupling to a readout SQUID current amplifier. A bias current $I_b$ together with a shunt resistor $R_{sh}$ provides a voltage bias across the TES. (b) Resistance and temperature relation of an Ir/Pt bilayer TES fabricated and measured at Argonne National Laboratory.}
	\label{fig:TES_diagram}
\end{figure}

A superconducting Transition-Edge Sensor (TES) photon detector, which utilizes a patterned superconducting film with a sharp superconducting-to-resistive transition profile as a thermometer, is a thermal detector with a well developed theoretical understanding. 
When a visible or infrared photon is absorbed by a TES, the tiny electromagnetic energy of the photon increases the temperature of the TES and therefore changes its resistance. 
See Figure~\ref{fig:TES_diagram}(b) for the temperature dependence of a TES resistance.
The change of the TES resistance is measured with a sensitive SQUID current amplifier in a voltage biased circuit as shown in Figure~\ref{fig:TES_diagram}(a), which provides linear operation and fast response because of a negative electro-thermal feedback~\cite{Irwin_05} to the detector.
TES photon detectors have been developed to measure single photons at a high efficiency for quantum communication~\cite{Rosenberg_07, Litaa_08, Fukuda_09, Fujino_11} and for axion-like particle searches in the shining light through the wall experiments~\cite{Bastidon_16, Eschweiler_15}. 
Applications also include large area photon measurement in direct detection of dark matter particles~\cite{Angloher_16, Rothe_18} for an enhanced background discrimination, calorimetric measurement of scintillation light and triplet excimers of superfluid helium~\cite{Hertel_19, Carter_17} for searching light dark matter particles, and high-precision astrophysical observations in the wavelengths between ultraviolet and infrared~\cite{Burney_06}. 
Moreover, TES detectors can be multiplexed enabling arrays of large channel counts~\cite{Bender_20, Doriesel_16, Dober_21}. 
Multiplexers for detector arrays using 16,000 TESs have already been successfully implemented~\cite{Bender_20}, and new developments exploiting microwave resonance techniques~\cite{Dober_21} have the potential to increase the multiplexing capacity by another factor of 10. 

A TES photon detector has an unparalleled energy resolution, which is an unique figure-of-merit advancing scientific discoveries. 
Operated under a negative electro-thermal feedback, the expected energy resolution~\cite{Irwin_05} of a TES detector is 
\begin{equation} 
\Delta E \approx 2.35 \sqrt{\frac{4k_B T_c^2 C}{\alpha} \sqrt{\frac{n}{2}}}, 
\label{eq:resolution} 
\end{equation} 
where $k_B$ is the Boltzmann constant, $T_c$ is the TES transition temperature, $C$ is the TES heat capacity, and $n$ is an index around five determined by electron-phonon decoupling~\cite{Giazotto_06}. 
Here $\alpha \approx (T/R)(dR/dT)$ is a parameter characterizing the TES transition profile, where $R$ is the temperature-dependent resistance of the TES in the transition. 
Therefore, a TES detector favors a low-$T_c$, a sharp superconducting-to-resistive transition with a large $\alpha$ and a small heat mass (through $C$).
A low-$T_c$ TES, which can be made of a W film~\cite{Abdelhameed_20, Fink_20} or an Ir-based bilayer~\cite{Yeomans_20}, is an effective way to improve energy detection resolution. 
Reducing the heat capacity of a TES by minimizing its size can also effectively enhance its resolution~\cite{Weii_08, Paoluccii_20}. 
An more advanced technology to reduce the heat capacity of a TES for a high resolution is to use Quasiparticle-trap-assisted Electro-thermal-feedback Transition-edge-sensor (QET)~\cite{Cabreraa_00, Saab_phd2002}. 
By utilizing a tailored TES with a small volume in intimate metallic contact with a large area superconducting photon absorber, a TES detector measures photons with a high resolution and with a large photon collection area.
As an enabling photon detection technology, the well-understood TES detector is a natural choice to count photons in many applications. 
There are several major R\&D and instrumentation opportunities summarized below:    

\begin{itemize}
    \item TES materials and fabrication techniques.  The researches include precise $T_c$ tuning in the range from sub-Kelvin down to 10 mK, QET technologies~\cite{Cabreraa_00, Saab_phd2002} for a large area photon detector with a high quantum efficiency, nanoscale TES detectors~\cite{Weii_08, Paoluccii_20}, and technologies to increase TES detector fabrication throughput. 

    \item Large area pixelated TES photon detector array for direct detection of dark matter particles. Each detector can be a patterned TES or a more advanced QET~\cite{Cabreraa_00, Saab_phd2002} with an enhanced photon collection area. The detector has a high energy resolution, therefore a low energy threshold covering the wavelengths from ultraviolet to infrared. 
    This large area high resolution photon detector is on demand in Sub-GeV dark matter searches.   
    The TESSERACT project, which is a R\&D project funded under DOE Dark Matter Small Projects New Initiatives~\cite{BRN_19}, requires such detectors to measure scintillation light and other excitations from detection targets including superfluid helium~\cite{Hertel_19}, gallium arsenide crystal~\cite{Derenzo_17} and sapphire crystal~\cite{Amare_05}.
    A large area high resolution photon detector is also required in new proposals searching for sub-Gev dark matter particles, such as high efficiency spectroscopic measurements of infrared photons from molecular vibration excitations of gases~\cite{Essig_19} and hydrogen-rich crystals~\cite{Wang_22} (which are water ice and hydrocarbons that have O-H and C-H bonds) for spin-dependent nuclear scatterings, and high efficiency and low threshold measurements of the fluorescence photons when a bosonic dark matter particle is absorbed by a molecule~\cite{Arvanitaki_18}. 
    Moreover, spectroscopic measurements with high resolution photon detectors help to identify a signal in searching for dark matter made out of axion-like particles or hidden photons. These include a dielectric haloscope~\cite{Baryakhtar_18} which converts axion-like particles and dark photons into photons in periodic photonic materials, and a spherical reflective surface dish antenna which converts axion-like particles and dark photons into photons emitted perpendicular to the surface~\cite{Horns_13}. 

    \item Ultra-sensitive TES photon detectors with a single sub-terahertz photon sensitivity~\cite{Weii_08, Paoluccii_20} for detection of dark matter axions at a large mass. The detectors are required to detect dark matter axions using a dielectric haloscope which consist of dielectric disks placed in a magnetic field to convert axions into photons~\cite{Caldwell_17}, or a haloscope that consists of a cylindrical metal barrel in magnetic field to convert axions into photons and a novel parabolic reflector to focus the photons onto a photon detector~\cite{Liu_21}.   
 
    \item TES infrared photon detectors with very high energy resolving power in astrophysics and cosmology. Such detectors are required not only in measurements of infrared radiations from ions, atoms, molecules, dust, water vapor and ice, but also astrophysical observations that trace our cosmic history from the formation of the first galaxies and the rise of metals to the development of habitable worlds and present-day life. The infrared detector researches are engineering challenges in TES sensor design, fabrication, and multiplexing readout of the large arrays~\cite{Nagler_18, Nagler_21}. 

    \item A TES detector with a large dynamic range and high quantum efficiency for quantum sensing. The photon number-resolving capability of the detector can allow a bit error probability that is unconditionally better than the standard quantum limit (SQL)~\cite{Thekkadath_21}.
\end{itemize}

\subsection{Novel Photoconductors for VUV photon detection} \label{sec:VUV}

Noble element TPCs produce a broad wavelength range of light from the vacuum ultraviolet (VUV) scintillation photons of ($<190$ nm), to ``blue'' Cherenkov photons (300nm - 600nm), and near infrared (NIR) light (600 - 2500nm). In a conventional noble element TPC the light detection devices utilize the semi-transparent charge readout planes (wires) and reside behind them. Although the noble elements themselves are highly transparent at these wavelength, the majority of commercially available optical detectors such as SiPMs and photomultiplier tubes are not sensitive to the VUV spectral range. While some devices now exist with VUV sensitivity \cite{Erdal:2016jhl,Igarashi:2015cma,Zabrodskii:2015pda}, this problem has traditionally been solved in large-scale systems by employing a wavelength shifting coatings to convert VUV light into a visible range where it can be detected by conventional sensors.  One of the most commonly used fluors is the organic compound tetraphenyl butadiene (TPB). However, the long term stability and behavior of TPB and other wavelength shifters is still largely unknown with some instabilities and undesirable effects already being noticed \cite{Asaadi:2018ixs}.

\subsubsection*{Amorphous Selenium}
Amorphous Selenium (a-Se) has long been identified as a useful photoconductor, first being widely used in photocopiers in the 1970's \cite{Weiss2017} and later garnering widespread commercial use in direct conversion active matrix flay panel imagers in the fields of digital mammography \cite{Kasap_2011} and digital breast tomosynthesis \cite{doi:10.1118/1.2903425}. The nature of amorphous materials means it easy to deposit over large areas (e.g. through thermal evaporation or similar techniques) without the use of expensive bonding processes. Moreover, a-Se detectors have a high quantum efficiency (QE), defined as $QE(E_\gamma) = 1 - e^{-\alpha d}$ where $\alpha$ is the absorption coefficent for a given energy of photon ($E_\gamma$) and $d$ is the thickness of the a-Se, in the X-ray regime where they have been typically used. The charge released ($Q_{\gamma}$) due to a photon interaction in the a-Se is characterized by $Q_\gamma = E_\gamma / W_\pm$ where $W_\pm$ is the conversion gain (in keV/electron-hole pair) and depends on a number of application specific parameters such as bias field, sensor thickness, and $E_\gamma$. For the energies in the X-Ray regime, a-Se based photon detectors have a very high QE and a good charge yield, which have made them very popular photon integrating imagers.

The usefulness of a-Se based detectors as photon counting devices has not been met with as much interest due to the relatively low charge gain when compared to other materials (such as silicon), low charge carrier mobility, and poor time resolution. This has held true until relatively recently, where amplification from avalanche gain due to impact ionization at high external fields allows a-Se based detectors to achieve charge conversion similar to crystalline semiconductors. These high-gain avalanche rushing photoconductors (HARPs) have been commercialized and used in the broadcast industry. Additionally, detector designs using multi-well solid state detectors have vastly improved the temporal response while simultaneously being able to achieve avalanche gain and through the introduction of various dopants (e.g. As, Cl, and CeO$_2$) the charge mobility has been improved while keeping the intrinsic dark current low ($\mathcal{O}$(pA/mm$^2$). 

A series of recent results have peaked the interest of the groups involved in this research enough to begin to explore the feasibility of using a-Se as a photoconductor in noble element TPC's. These results include an a-Se p-n junction device sensitive to UV light \cite{doi:10.1063/1.3579262}, microfabricaiton of a UV sensitive a-Se detector \cite{article:MicroASe}, the production of a hybrid a-Se CMOS photon counting sensor \cite{inproceedings:aSeCMOS}, and timing resolutions less than 1 nanosecond for a-Se avalanche detectors \cite{doi:10.1021/acsphotonics.9b00012}. Taken together, it seems that the synthesis of these approaches along with integration into a noble element TPC would provide a potentially game changing method for direct UV photon detection with the possibility to expand to a wider range of wavelengths. 

\begin{figure}[htb]
    \centering
    \includegraphics[width=0.45\textwidth]{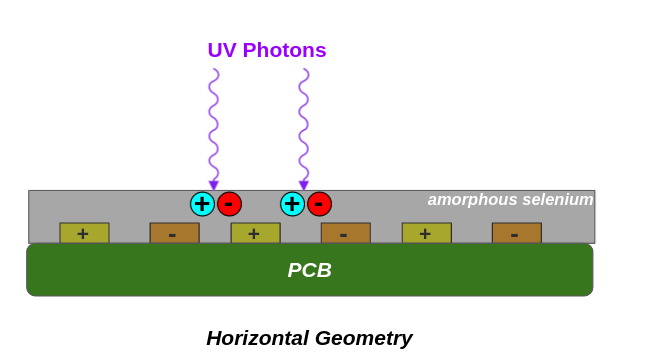}
    \includegraphics[width=0.45\textwidth]{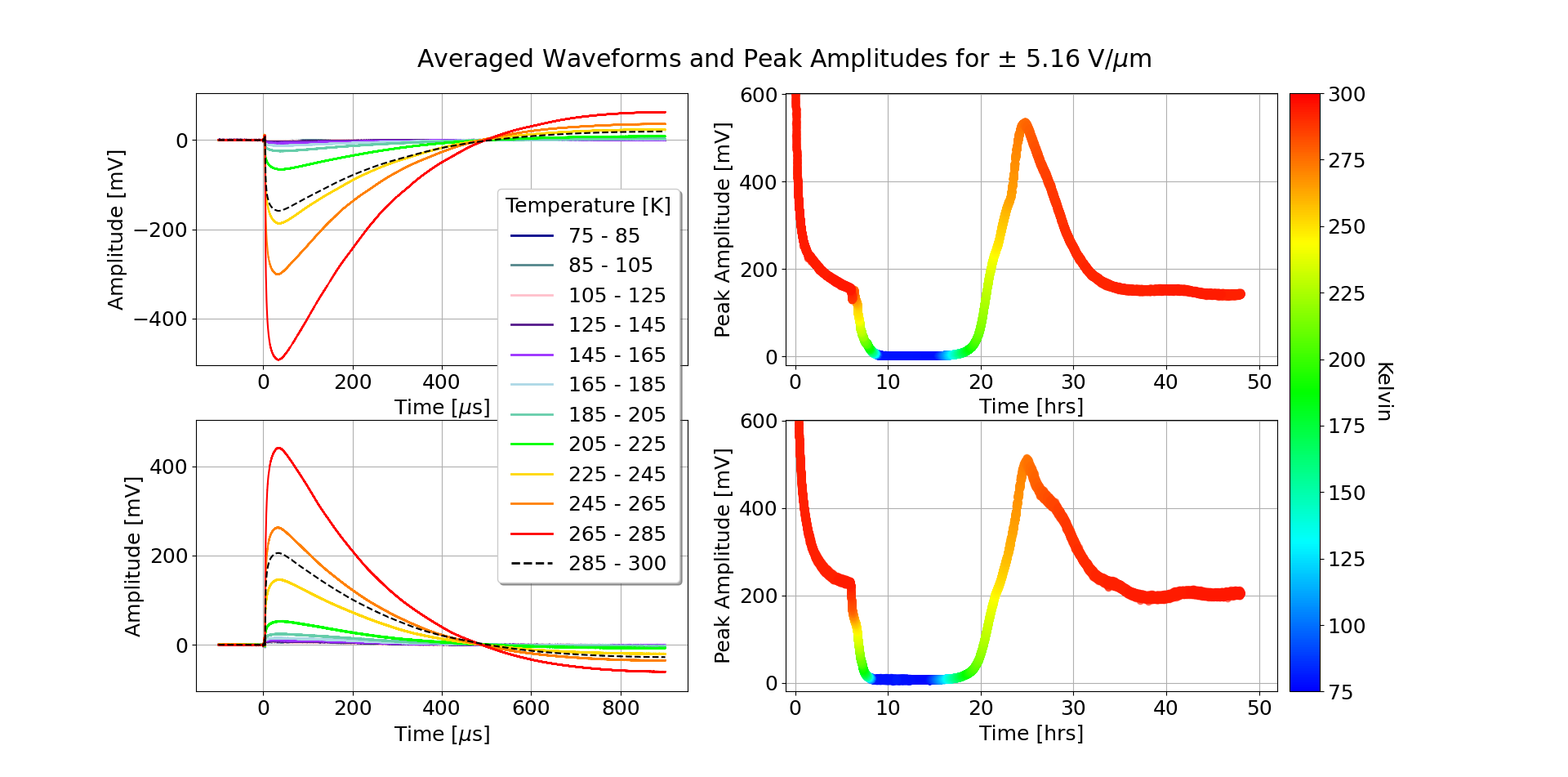}
    \caption{Left: Schematic the horizontal geometry under development for absorption of UV light by an aSe based detector. Right: Early results showing the response of an aSe based photon detector as a function of temperature ranging between 290~K to 77~K.}
    \label{fig:aSe}
\end{figure}

Recent experimental work has shown the viability of aSe in a cryogenic environment to respond to VUV light at low applied electric field ($E \leq 5$V/$\mu$m). This technique utilizes a non-standard geometry for aSe detectors dubbed a ``horizontal'' configuration in order to circumvent the fact that the typical electrodes used (e.g. ITO, gold, copper, etc) will result in a large fraction of all the UV light being absorbed. A ``horizontal geometry'' can be constructed from a bare printed circuit board (PCB) with interdigitated electrodes, as shown in left of Figure \ref{fig:aSe}. The selenium is thermally evaporated directly onto the board and thus can be exposed directly to the UV source. The right of Figure \ref{fig:aSe} shows an example of the response of such a aSe detector to VUV light across various temperatures. While the size of the signal does decrease as a function of temperature, the detector does continue to respond all the way down to $\sim 77$K. 

Additional work is ongoing to adapt this concept to allow for higher applied electric fields $\mathcal{O}$(80~V/$\mu$m) in order to allow charge gain and to explore the design considerations needed for single photon sensitivity. 

\subsubsection*{Organic Semiconductors}
The field of organic semiconductor devices has seen rapid growth in recent years.  Their use can be found in a wide range of devices from QLED displays used in cell phones to quantum-dot based televisions, to wearable tech, and in the form of photovoltaics for solar energy.  They are relatively simple and cheap to produce, easily adaptable for use with modern additive manufacturing technologies, and are highly tunable with regards to their optical bandgap and chemical structure.  Recent work on organic photodiodes has demonstrated their low noise can be on the level of state-of-the-art silicon photodiodes \cite{Fuentes-Hernandez2020}. These properties offer exciting possibilities for their use high energy physics experiments.  Possible applications include, large-area pixeled photosensors which could be additivity manufactured on rigid or flexible substrates.  The use of VUV transparent conductive electrodes would allow for direct detection of VUV scintillation light in liquid noble gas detectors.  Due to their highly tunable optical bandgaps, a stack of organic photodiodes can be used as a spectroscopic or color sensitive photosensor. This could provide for example, separation of Cherenkov and scintillation signals within the same device without loss of photo coverage.  This type of stack geometry has already been demonstrated for red-green-blue sensitive optical sensors \cite{Seo2007}.

While offering numerous advantages there are still significant R\&D technical challenges which need to be addressed.  The nature of charge transport within $\pi$-conjugated materials results in a slower response time compared to silicon photosensors \cite{Coropceanu2007}. While high photoconductor gain has been demonstrated for CW light \cite{Chen2008}, gain has yet to be demonstrated for faint pulsed light characteristic of scintillators.  Their performance in extreme environments such as cryogenic temperature or high radiation fields. Investigations into possible VUV transparent electrodes are warranted for use in direct detection of liquid noble gas detector VUV scintillation light.


\section{Summary}

 In Section~\ref{sec:needs} we discussed the needs for photon counting developments in the next generation of HEP experiments. The next generation of spectroscopic cosmic surveys will become more efficient with photon counting sensors and improved sensitivity in the IR (see Sec.~\ref{sec:cosmology}). These technologies could also play an important role in future space telescope missions. Photon counting technologies have also demonstrated world leading performance in the direct search for dark matter (Sec.~\ref{sec:darkmatter}) and low energy neutrino experiments (Sec.~\ref{sec:neutrinos}). Furthermore, unlocking the full potential of noble elements TPCs requires large area, high UV efficiency sensors capable of detecting both light and charge. These developments needed for HEP will also impact quantum sensing, research in basic energy science and low energy radiation detection applications as discussed in Sec.~\ref{sec:other}.
 
 In Section~\ref{sec:techs} we discussed the opportunity to address these needs with technology currently being developed as part of the HEP program. 
 Photon counting in the visible can be achieved now with semiconductor technologies with large number of pixels in skipper-CCDs and CMOS sensors (Sections~\ref{sec:skipper} and~\ref{sec:cmos} ), the capabilities of these sensors are being extended into lower energies with Ge as discussed in Sec.~\ref{sec:gedet}. 
 SiPMs allow single photon resolution in the visible range and will be assembled into very large area photon detection systems, in particular for noble liquid detectors.
Research is pursued to enhance their sensitivity both at NIR and VUV wavelengths (Sec.~\ref{sec:SPAD_VUV} and~\ref{sec:SPAD_NIR}).
PDCs perform the direction conversion of photon counts to digital signal and offer low power consumption, low time jitter capability, and embedded signal processing (Sec.~\ref{sec:PDC}).
 Photon counting with higher time resolution and for lower energy photons is possible in superconducting detectors TES (Sec.~\ref{sec:TES}), MKIDs (Sec.~\ref{sec:MKID}) and SNSPDs (Sec.~\ref{sec:SNSPD}), with significant effort ongoing to make larger arrays of these cryogenic detectors. 
 Novel photon detectors for VUV are being developed for noble elements TPCs as discussed in Sec.~\ref{sec:VUV}. 
 
 The photon counting technologies presented here are maturing fast. The further development of these promising photon counting technologies as part of HEP is expected to have a large impact on the field in the coming decade.







\bibliographystyle{plain}
\bibliography{bibliography.bib,mazin3.bib,UdeS_references_1.bib,UdeS_references_2.bib,DBraga.bib}









\end{document}